\documentclass{jfm}

\usepackage{graphicx}
\graphicspath{{./Figs/}}
\usepackage{epstopdf, epsfig}
\usepackage{newtxtext}
\usepackage{newtxmath}
\usepackage{natbib}
\usepackage[caption=false]{subfig}
\newcommand{\RomanNumeralCaps}[1]
\linenumbers

\usepackage{hyperref}
\hypersetup{
  colorlinks = true,
  urlcolor   = blue,
  citecolor  = black,
}
\usepackage{amsmath}
\usepackage{color}
\usepackage{comment}

\newcommand{\ba}{\boldsymbol{a}}
\newcommand{\bx}{{\boldsymbol{x}}}
\newcommand{\br}{{\boldsymbol{r}}}
\newcommand{\bu}{{\boldsymbol{u}}}

\newcommand{\avg}[1]{\langle #1 \rangle}
\newcommand{\cg}[1]{\langle #1 \rangle_\ell}
\newcommand{\epsl}{\varepsilon_\ell}
\newcommand{\Bl}{\mathcal{B}_\ell}
\newcommand{\rhop}{\rho_\mathrm{p}}
\newcommand{\taup}{\tau_\mathrm{p}}
\newcommand{\xp}{\boldsymbol{x}_\mathrm{p}}
\newcommand{\vp}{\boldsymbol{v}_\mathrm{p}}
\newcommand{\ap}{\boldsymbol{a}_\mathrm{p}}

\newcommand{\St}{\textit{St}}
\newcommand{\Stl}{\St_\ell}
\newcommand{\Rel}{\Rey_\ell}
\newcommand{\Pel}{\Pen_\ell}

\title{Homogeneous turbophoresis of heavy inertial particles in turbulent flow}

\author[J. Bec \& R. Vall\'{e}e]{J\'{e}r\'{e}mie Bec\aff{1,2}\corresp{\email{jeremie.bec@univ-cotedazur.fr}} and Robin Vall\'{e}e\aff{3}}
\affiliation{\aff{1}Universit\'{e} C\^{o}te d'Azur, CNRS, Institut de Physique de Nice, France \aff{2}Universit\'{e} C\^{o}te d'Azur, Inria, CNRS, Calisto team, Sophia Antipolis, France \aff{3}Mines Paris, PSL University, CNRS, Cemef, Sophia Antipolis, France}

\begin{document}

\maketitle

\begin{abstract}
  Particles suspended in turbulent flows are commonly found in nature and industry, appearing as droplets, dust, or sediments. When heavier than the fluid, they possess inertia and are ejected from the most violent vortical structures of the carrier flow by centrifugal forces. Once piled up along particle paths, this small-scale mechanism leads to an effective large-scale drift. This phenomenon, known as ``turbophoresis,'' causes particles to leave highly turbulent regions and migrate towards calmer regions, resulting in uneven spatial distributions. This process has been extensively studied and is used to explain why particles transported by non-homogeneous flows tend to concentrate near the minima of turbulent kinetic energy.

  It is demonstrated here that turbophoretic effects are just as crucial in statistically homogeneous flows. Although the average turbulent activity is uniform, instantaneous spatial fluctuations trigger local fluxes that are responsible for inertial-range inhomogeneities in the particle distribution.  Direct numerical simulations are used to thoroughly probe and depict the statistics of particle accelerations, specifically their scale-averaged properties conditioned on local turbulent activity.  The simulations confirm the relevance of the local energy dissipation to gauge instantaneous spatial fluctuations of turbulence. This analysis yields an effective coarse-grained dynamics, which accounts for particle detachment from the fluid and their ejection from excited regions through a space and time-dependent non-Fickian diffusion.

  Such considerations lead to cast fluctuations in particle distributions in terms of a scale-dependent P{\'{e}}clet number $\mbox{\textit{Pe}}_\ell$, which measures the relative importance of turbulent advection compared to inertial turbophoresis at a given observation scale $\ell$. Multifractal statistics of the coarse-grained turbulent energy dissipation indicate that $\mbox{\textit{Pe}}_\ell \sim \ell^\delta/\tau_\mathrm{p}$ with $\delta\approx 0.84$.  Numerical simulations support this behaviour and emphasises the relevance of the turbophoretic P{\'{e}}clet number in characterising how particle spatial distributions, including their radial distribution function, depends on $\ell$. This approach also explains the presence of voids with inertial-range sizes, and the fact that their volumes has a non-trivial distribution with a power-law tail $p(\mathcal{V}) \propto \mathcal{V}^{-\alpha}$, with an exponent $\alpha$ that tends to 2 as $\mbox{\textit{Pe}}_\ell\to0$.  In addition to its ability to describe particle concentrations, the proposed approach provides a new framework for modelling particle transport in large-eddy simulations of turbulent flows.

\end{abstract}

\begin{keywords}
  Particle/fluid flow, Isotropic turbulence, Intermittency
\end{keywords}


\section{Introduction}
\label{sec:intro}

The transport of small, heavy particles by a developed turbulent flow is a common occurrence in nature and industry. Whether they are droplets in air, dust in gas, or sediments in water, these particles are often smaller than the smallest active scale of the fluid and have a larger mass density. They thus possess inertia, resulting in their detachment from the carrier fluid and uneven spatial distributions, a phenomenon known as \textit{preferential concentration}. This is important in determining the interactions between these particles, such as collisions and aggregation. It also alters the transfers of momentum, kinetic energy, and heat in the particle-laden fluid.  One notable example of inertial particles is  water droplets in atmospheric clouds. As stressed by \cite{jonas1996turbulence}, turbulence triggers variability in droplet sizes that can explain why the timescales for rain initiation are much shorter than those predicted by mean-field arguments. \citet[][see also \citealt{shaw2003clouds}]{pinsky1997turbulence} demonstrated that the preferential concentration of droplets affects their growth by condensation and coalescence. Heterogeneities have been observed in situ \citep[see, \textit{e.g.},][]{kostinski2001droplet} and their small-scale effects have been quantified to improve droplet collision rates \citep[see][]{reade2000numerical,falkovich2002droplet}.  Still, many challenging questions raised in clouds involve interactions over a huge range of scales and thus, cannot be addressed without having recourse to large-eddy simulations (LES).  Such approaches need \textit{ad-hoc} parameterisations of particle dynamics and their microphysical interactions, as discussed for instance in \cite{morrison2020confronting}.  Planet formation by dust aggregation in the early Solar system is another important natural instance of inertial particles, which raises similar issues. Local fluctuations in the particle concentration trigger gravitational collapse and thus the formation of larger objects.  Because of rotation around the star, dust particles migrate in large-scale anticyclonic Keplerian vortices \citep{gerosa2023clusters} or in pressure bumps \citep{johansen2007rapid}.  It is probably in these regions that primary accretions occur, but the effect of turbulence is still unclear \citep{johansen2015new}.  A better understanding requires developing models to quantify dust clustering in the inertial range of turbulence \citep[see, \textit{e.g.},][]{hartlep2020cascade} and designing LES tools that cope with astrophysical specificities. Other natural situations where inertial particles occur include plankton ecology in the ocean \citep{seuront2001turbulence} and seed dispersion above plant canopies \citep{pan2014canopy}. In all cases, a precise description of large-scale fluctuations in particle density is crucial.

Equivalent questions arise in engineering. When optimising droplet vaporisation in injection sprays \citep{sahu2018interaction} or monitoring particulate fouling \citep{henry2012towards}, it is important to understand how inertial particles distribute over scales comparable to the larger scales of the carrier turbulent flow.  The complexity of flow geometries and  inhomogeneities in industrial applications give a critical role to the spatial variations of the time-averaged particle density. Much effort has thus been dedicated to derive effective transport equations for the average concentration field. In this context, \cite{caporaloni1975transfer} unveiled a fundamental mechanism in which turbulence inhomogeneities drive particles out of the most excited regions of the flow and concentrate them in quieter zones.  They dubbed this phenomenon \textit{turbophoresis} \citep[see also][]{reeks1983transport}, in analogy to thermophoresis, where temperature gradients cause a motion of diffusive particles toward colder regions of space. \citet{reeks1983transport,reeks1992continuum} proposed closures of the kinetic equations for the particle phase-space distribution to derive effective diffusion equations for the average spatial concentration. This leads to the particle fluxes due to inertia being described by a Fick law, where the coefficient of diffusion is related to the local Lagrangian correlation of the fluid velocity. Such arguments have been successfully employed to explain why particles in turbulent channel flows tend to migrate towards the walls \citep[see, \textit{e.g.},][]{marchioli2002mechanisms,kuerten2005can,sardina2012wall,fouxon2018inhomogeneous,brandt2022particle}.  However the dependence of the diffusion coefficient on the particle Stokes number is not yet fully understood. \citet[see also \citealt{belan2016concentration}]{belan2014localization} showed that particles with sufficient inertia escape from low-kinetic-energy regions, leading to a localisation/delocalisation phase transition. \cite{delillo2016clustering} examined the case of turbulent flows with an inhomogeneous forcing and found that turbophoretic effects are more pronounced at intermediate particle inertia. \cite{mitra2018turbophoresis}  interpreted this behaviour as a balance between turbophoretic and turbulent diffusions. 

The applicability of turbophoresis to particle transport in flows with average inhomogeneities raises questions about its relevance in homogeneous situations.  In homogeneous isotropic turbulence, instantaneous snapshots reveal spatial fluctuations of kinetic energy throughout the inertial range.  Meanwhile, particle distributions display heterogeneous concentrations characterised by large-scale quasi-uniform regions, localised voids, and sheet-like clusters, as observed for instance by \cite{eaton1994preferential}.  To quantify inertial-range particle distributions, different observables are needed compared to those used for the dissipative range.  At small scales, particle distributions exhibit multifractal scaling properties \citep[see][]{hogan1999scaling,bec2011spatial,schmidt2017inertial,bec2024statistical} and are fully characterised by a dimension spectrum that depends solely on the Stokes number.  The unified picture of the joint dependence on length scale and response time arises from the fact that dissipative-range dynamics involve a unique timescale determined by the typical amplitude of velocity gradients. This is in contrast with the hierarchy of timescales involved in inertial-range physics. In the two-dimensional inverse cascade, \cite{boffetta2004large} found that particles concentrate quasi uniformly on thin filamentary structures separated by voids whose distribution follows a universal scaling law.  However, in the random, white-in-time, self-similar flows considered by \cite{bec2007clustering}, such scaling is  absent, and particle distributions are characterised by local fractal dimensions determined by the scale-dependent Stokes number $\Stl \propto \taup/\ell^{2/3}$ \citep{balkovsky2001intermittent}, defined by non-dimensionalising the particle response time by the turnover time at the observation scale $\ell$. Both of these scenarios coexist in three-dimensional turbulence, as pointed up by \cite{bec2007heavy}, \cite{yoshimoto2007self}, or inferred from the \textit{sweep-stick} mechanism of \cite{goto2008sweep}. The intricate spatial correlations of the pressure gradient, or equivalently of the fluid acceleration, play a key role. By using Vorono\"{\i} tesselations, \cite{monchaux2010preferential,monchaux2012analyzing} introduced a definition of particle clusters and found that their size distribution follows a universal scaling law independent of the Stokes number. This was confirmed by \cite{baker2017coherent}, who showed that clusters preferentially sample regions of the flow with higher strain and lower vorticity. However, \cite{bragg2015mechanisms} found that this statistical bias depends on inertia and is actually quantified by the scale-dependent Stokes number $\Stl$. These arguments led them to predict scale invariance for two-particle statistics when $\Stl\ll 1$, which was confirmed by \cite{hartlep2017scale} using a cascade multiplier approach.  \citet{ariki2018scale} further argued that the pair correlation function follows a universal power-law $\propto \Stl^2$ using a Lagrangian renormalisation closure.  The wavelet analysis conducted by \citet{matsuda2021scale} shows intermittent particle densities, with a stronger contribution from voids observed at smaller spatial scales. However, the question of whether scale invariance holds in the inertial-range distributions of particles and, if so, which mechanisms are involved, remains ambiguous.

To shed new lights on these issues, an appropriate effective model for inertial-range particle dynamics is expected to be useful. While there are various simplified approaches to dilute particle suspensions, reviewed for instance by \cite{balachandar2010turbulent}, the Eulerian field representations of the particle phase proposed by \cite{ferry2001fast} provide promising tools. In this approach, the particle velocity is enslaved to the carrier phase, with the effect of inertia being regarded as a compressible correction proportional to the fluid velocity acceleration.  While this approximation has often shown its relevance, it remains limited to the asymptotics of small particle inertia, and it combines very different timescales, as acceleration is influenced by dissipative-range physics.  \cite{fevrier2005partitioning} extended these considerations to large Stokes numbers by assuming that the particle motion can be seen as the sum of a mesoscopic velocity and a random component.  The latter term corresponds to a diffusive motion, is uncorrelated in space, and has been found to properly reproduce particle properties when their response time is much larger than the turbulent large-eddy turnover time. This contribution, dubbed \textit{random uncorrelated motion} by \cite{reeks2006search},  was used by \cite{gustavsson2012inertial} in synthetic random flows and shown to suitably describe the effect of fold caustics on the particle kinetics.  This approach relies on the idea that turbulence has only a cumulative effect along particle paths, as long as the latter have a sufficiently long correlation time. However, fluctuations do not need to be averaged over times prescribed by the particles lag, but this procedure can rather stem from a spatial or temporal coarse-graining of the turbulent field, thus incorporating the effect of instantaneous spatial inhomogeneities.

We aim here to present a model that can effectively describe and quantify particle dynamics in the inertial range of a fully developed turbulent flow.  Small-scale detachments from the fluid result in particles carrying forward past fluctuations instead of filtering them out in a time-reversible manner.  Building on the phenomenology introduced by \cite{bec2007toward}, we argue that this mechanism cumulates over time, leading to an ejection process that causes non-Fickian particle fluxes.  Our proposed model utilizes an It\^o, rather than Stratonovich, diffusion process with a diffusion coefficient that varies based on the local flow activity. This statistically homogeneous turbophoresis can be used to quantify inhomogeneities in the particle distribution and, to some extent, reconcile the various viewpoints discussed above. Our analysis is grounded in the results of direct numerical simulations conducted at large Reynolds numbers and relies on a comprehensive evaluation of particle accelerations. 

The paper is structured as follows.  In \S\ref{sec:model} we introduce our settings and discuss the relevant observables for our analysis. We also provide a general appreciation of the correlations between particle concentrations and instantaneous inhomogeneities in turbulent activity. In \S\ref{sec:lagrangian} we develop a Lagrangian perspective and conduct a detailed statistical analysis of particle acceleration. In \S\ref{sec:eulerian} we shift our focus to the Eulerian frame and derive an effective equation for the particle coarse-grained density. From this model, we draw properties of the inertial-range distribution and discuss specifically the implications of this approach to the distribution of voids. Finally, in \S\ref{sec:conclusion} we summarise our findings and discuss possible perspectives.


\section{Models, simulations, and spatial coarse-graining}
\label{sec:model}

\subsection{Homogeneous isotropic turbulence and energy dissipation}
\label{subsec:turbu}

We investigate the behaviour of particles passively suspended in a three-dimensional fluid flow. The velocity field of the fluid, denoted by $\bu(\boldsymbol{x},t)$, satisfies the incompressible Navier--Stokes equations
\begin{equation}
  \partial_t \bu + \bu \boldsymbol{\cdot \nabla} \bu =
  - (1/\rho_\mathrm{f})\,\boldsymbol{\nabla} p + \nu \nabla^2 \bu +\boldsymbol{f} \quad \text{with} \quad \boldsymbol{\nabla\cdot} \bu = 0,
  \label{eq:NS}
\end{equation}
where $p$ represents the pressure, $\rho_\mathrm{f}$ is the mass density of the fluid, $\nu$ is its kinematic viscosity, and $\boldsymbol{f}$ is an external volume force. The force is prescribed with homogeneous and isotropic statistics and is correlated on large scales in both space and time. The force injects kinetic energy into the flow at an average rate of $\varepsilon = \langle\boldsymbol{f\cdot}\bu\rangle$. We perform direct numerical simulations of (\ref{eq:NS}) using the pseudo-spectral code \textit{LaTu} on the triply periodic box $[0,2 \pi]^3$ and employ third-order Runge--Kutta time marching.  The details of the code can be found in \cite{homann2007impact}. Two sets of simulations are carried out with different resolutions. Corresponding numerical and turbulent parameters are presented in table~\ref{tab:param}. 
\begin{table}
  \begin{center}
\def~{\hphantom{0}}
  \begin{tabular}{ccccccccc}
    $N^3$  & $\nu$   &   $\Delta t$ & $\varepsilon$ & $\eta$ & $u_\mathrm{rms}$ & $L$ & ~$R_\lambda$~ & $N_\mathrm{p}$ 
    \\ \hline
    $1024^3$   & ~\,\,$6\times10^{-5}$ & ~$0.003$& $3.47\times10^{-3}$ & $2.81\times10^{-3}$ & $0.185$ & $1.82$ & $290$ & $1.25\,10^{7}$\\
    $2048^3$   & $2.5\times10^{-5}$ & $0.0012$ & $3.61\times10^{-3}$ & $1.44\times10^{-3}$ &$0.189$ & $1.87$ & $460$ & ~~~~$10^{8}$\\
  \end{tabular}
  \caption{Parameters of the numerical simulations: $N^3$ number of collocation points; $\nu$~fluid kinematic viscosity; $\Delta t$ time step; $\varepsilon$ average kinetic-energy dissipation rate; $\eta=\nu^{3/4}/\varepsilon^{1/4}$ Kolmogorov dissipative scale; $u_\mathrm{rms}$ root-mean-square velocity; $L=u_\mathrm{rms}^3/\varepsilon$ large scale; $R_\lambda = \sqrt{15}\,u_\mathrm{rms}^2/(\varepsilon\,\nu)^{1/2}$ Taylor-scale Reynolds number; $N_\mathrm{p}$ number of particles for each value of the Stokes number.}
  \label{tab:param}
  \end{center}
\end{table}

After a certain period of time, the fluid velocity field $\bu$ reaches a statistical steady state characterised by multifractal statistics of the local dissipation rate $\varepsilon_\mathrm{loc}(\bx) = (\nu/2)\,\mathrm{tr}\,(\boldsymbol{\nabla}\bu(\bx)+\boldsymbol{\nabla}\bu^\mathsf{T}(\bx))^2$ \citep[see, \textit{e.g.},][]{frisch1995turbulence}. This is evidenced from the scale-dependent statistics of the coarse-grained dissipation $\epsl$ obtained by averaging the local dissipation over the ball $\Bl(\bx)$ of center $\bx$ and diameter $\ell$, that is  $\epsl(\bx) \equiv (1/|\Bl|)\int_{\Bl(\bx)}\varepsilon_\mathrm{loc}(\bx')\,d^3x'$. For $\eta\ll\ell\ll L$, the probability distribution of $\epsl$ takes the form
\begin{equation}
  p(\epsl)\,\mathrm{d}\epsl = \left(\ell/L \right)^{3-\mathcal{D}(\alpha)}\mathrm{d}\mu(\alpha) \quad \mbox{with}\quad \epsl = \varepsilon\,(\ell/L)^{\alpha},
  \label{eq:multif}
\end{equation}
where $\mathcal{D}(\alpha)$ is the multifractal spectrum, which can be interpreted as the dimension of the fractal set on which the scale-averaged dissipation is $\propto \ell^{\alpha}$ when $\ell/L\to0$, and $\mathrm{d}\mu(\alpha)$ corresponds to the weight associated with each singularity exponent $\alpha$.  In Kolmogorov 1941 phenomenology, there are no fluctuations of $\epsl$ and  $\mathcal{D}(\alpha)=-\infty$ except for $\alpha=0$, for which $\mathcal{D}(0) = 3$.
\begin{figure}
\centering
\subfloat{\includegraphics[width=0.49\columnwidth]{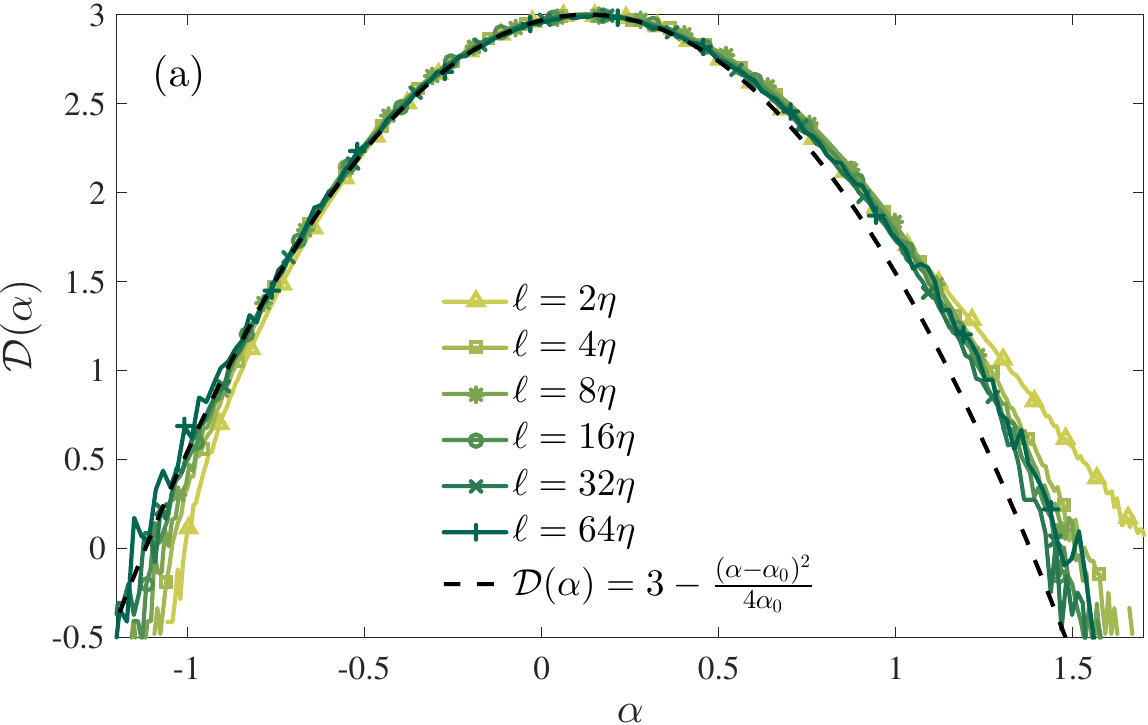}\label{fig:lognormal}}
\hfill
\subfloat{\includegraphics[width=0.49\columnwidth]{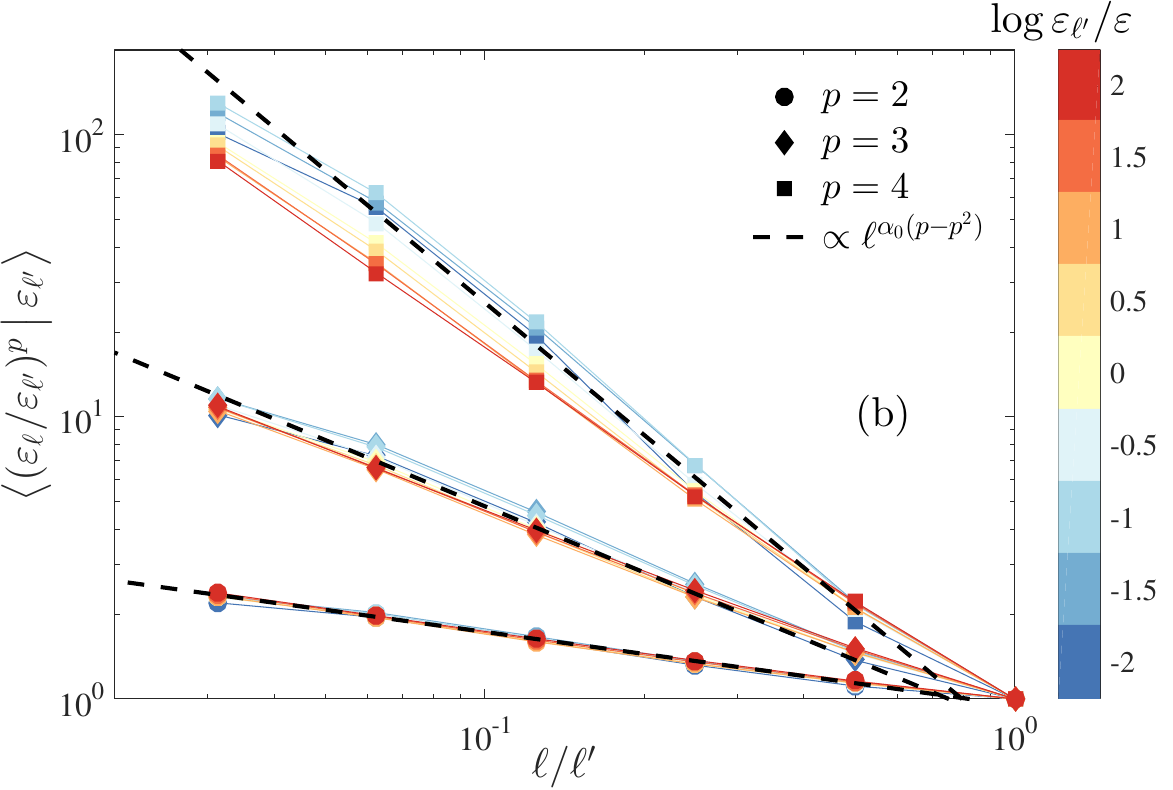}\label{fig:stat_epsil_cond}}
\caption{Statistics of the scale-averaged dissipation rate $\epsl$ for $R_\lambda = 460$. (a) Measured dimension spectrum $\mathcal{D}(\alpha)$ as a function of $\alpha = \log(\epsl/\varepsilon) / \log(\ell/L)$ at various scales $\ell$. The lognormal approximation for $\alpha_0=0.26$ is shown as a dashed line. (b) Moments of order $p=2$, $3$, $4$ of the average dissipation rate $\epsl$ conditioned on its value $\varepsilon_{\ell'}$ in a larger box of size $\ell^\prime=128\,\eta$, as a function of $\ell/\ell^\prime$ and different conditionings as indicated in the colour bar. Lognormal behaviours are displayed as dashed lines.}
\end{figure}
Figure~\ref{fig:lognormal} shows the multifractal spectrum obtained from numerical measurements of $\epsl$. We find that lognormal statistics, for which the dimension spectrum is a parabola $\mathcal{D}(\alpha) = 3-(\alpha-\alpha_0)^2/(4\alpha_0)$, provide a good approximation for the lowest values of $\alpha$. It is however known that lognormal distributions have several shortcomings due to the non-conservative nature of the cascade models on which they are based \citep[see discussion in][\S8.6.5]{frisch1995turbulence}. Despite this, such an approximation is still useful for estimating moderate-order statistics, well beyond the central-limit approximation.  Using \cite{kolmogorov1962refinement} refined similarity hypothesis, and recent confirmations by \cite{lawson2019direct}, the statistics of the fluid velocity relate to the fluctuations of $\epsl$.  For the longitudinal structure functions $S_n^\parallel (\ell) = \left\langle \boldsymbol{[\hat{\ell}\cdot} (\bu(\bx+\boldsymbol{\ell})-\bu(\bx))]^n \right\rangle$, the lognormal approximation with parameter $\alpha_0$ predicts a scaling behaviour  $S_n^\parallel (\ell)  \sim \ell^{\zeta_n}$ with  $\zeta_n = n/3+(\alpha_0/9)(3n-n^2)$.  For $\alpha_0=0.13$ obtained from our simulations, we get $\zeta_2\approx 0.696$, $\zeta_4\approx 1.276$, $\zeta_6\approx 1.740$, which are in good agreement with experimentally measured values \citep[see][for a recent review]{saw2018universality}.

Multifractal statistics are often interpreted phenomenologically as resulting from the random multiplicative cascade experienced by the coarse-grained dissipation. This scenario suggests that the probability distribution (\ref{eq:multif}) should also apply to the fluctuations of $\varepsilon_{\ell}(\bx)$ conditioned on the observed value of $\varepsilon_{\ell^\prime}(\bx)$ at the same location but over a larger scale $\ell^\prime>\ell$. As shown in figure~\ref{fig:stat_epsil_cond} for $\ell^\prime = 128\eta$, numerical simulations confirm this feature, revealing a scaling regime with an exponent that is closely approximated by the lognormal prediction.  These multiscale statistics play a crucial role in investigating the coarse-grained dynamics of transported particles, as we will discuss in more detail later on.

\subsection{Particles, preferential sampling and concentrations}
\label{subsec:concentrations}

After the fluid flow reaches a statistical steady state, we introduce heavy, inertial, point-like particles that are homogeneously seeded with velocities equal to that of the fluid at their positions. The trajectories $\xp(t)$ of these particles follow
\begin{equation}
  \frac{\mathrm{d}\xp}{\mathrm{d}t} = \vp,\quad \frac{\mathrm{d}\vp}{\mathrm{d}t} = \ap = -\frac{1}{\taup}\left[\vp - \bu(\xp,t)\right],
  \label{eq:viscous_drag}
\end{equation}
Particles are assumed much smaller than the Kolmogorov dissipative scale $\eta$, and sufficiently massive to neglect so added-mass, Magnus, and history effects. The viscous drag intensity is given by the response time $\tau_{\rm p} = \rho_{\rm p}\,d_{\rm p}^2 / (18\,\nu\,\rho_{\rm f})$, where $\rho_{\rm p}$ is the particle mass density and $d_{\rm p}$ its diameter.  This time is used to define the Stokes number $\St = \tau_{\rm p} / \tau_{\rm \eta}$, with $\tau_{\rm \eta} = (\nu / \epsilon)^{1/2}$ denoting Kolmogorov dissipative timescale. The Stokes number measures particle inertia. When $\St\ll 1$, the particles almost follow the flow and behave as tracers. When $\St\gg 1$, they detach from the flow and behave ballistically. We adopt a Lagrangian approach in our simulations, where particles trajectories are tracked by integrating Eq.~(\ref{eq:viscous_drag}) with the fluid velocity at their location obtained by linear interpolation from the grid. We use 10 different values of the Stokes number in the range $\St \in [0.1, 6.5] $ and, for each value of $\St$, a number $N_\mathrm{p}$ of particles that roughly corresponds to one particle per box of size $(9\eta)^3$.

\begin{figure}
  \centering
  \subfloat{\includegraphics[width=\columnwidth]{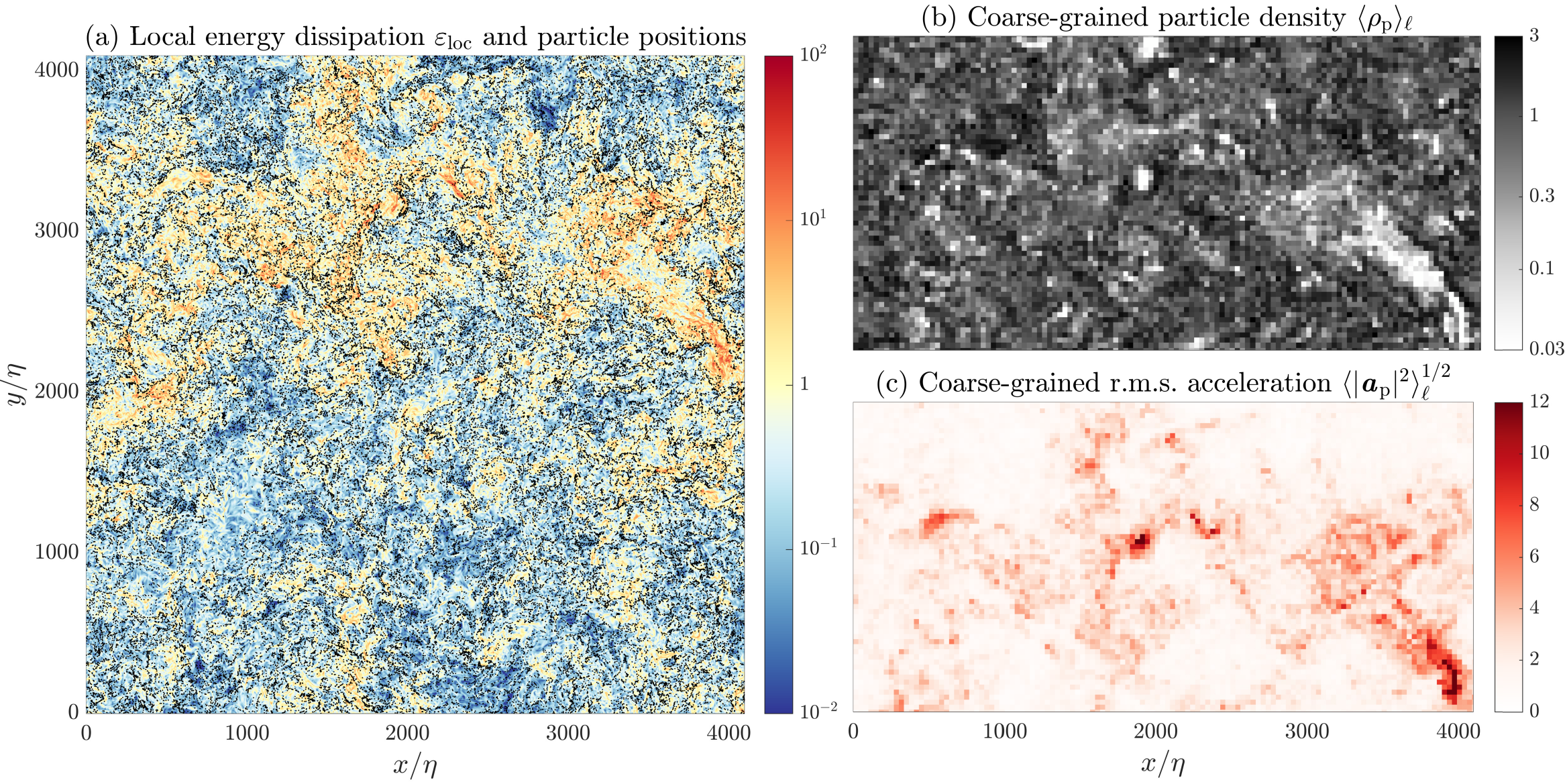}\label{fig:snap_epsil_part_a}}
  \subfloat{\label{fig:snap_epsil_part_b}}
  \subfloat{\label{fig:snap_epsil_part_c}}
\caption{(a) 2D slice of the instantaneous energy dissipation field $\varepsilon_\mathrm{loc}(\bx)$ for $R_\lambda=460$, together with particle positions for $\St=1$ shown as black dots. Colours stand for $\log_{10}(\varepsilon_\mathrm{loc} /\varepsilon)$.  (b) Particle coarse-grained density $\cg{\rhop}$ in the upper half of the same slice obtained for $\ell = 32\,\eta$. Colours are again on a logarithmic scale. (c) Root-mean square particle acceleration, coarse-grained over the same grid, here normalised by $(\varepsilon^3/\nu)^{1/4}$.}
  \label{fig:snap_epsil_part}
\end{figure}
Upon reaching a statistically stationary state, the particle distributions exhibit highly non-uniform patterns and strongly correlate with the turbulent structures of the flow, as depicted in figure~\ref{fig:snap_epsil_part_a}.  The spatial arrangement of particles shows voids in the most active regions of the flow, where dissipation is high, sheet-like clusters that encapsulate these voids, and quasi-uniform distributions in regions with lower turbulent intensity. These concentration fluctuations are attributed to the inertial-range motions of particles, as the sizes of the regions are much larger than the dissipative scale $\eta$. To filter out dissipative-range effects, we introduce the coarse-grained particle density $\cg{\rhop}$. It is obtained by counting the number of particles in small boxes of size $\ell$, which define a partition of the spatial domain.  Figure~\ref{fig:snap_epsil_part_b} shows $\cg{\rhop}$ obtained with  $\ell = 32\,\eta$. The spatial variations of  particle dynamics also serve as a marker for the different regions of the flow. In figure~\ref{fig:snap_epsil_part_c}, we show the coarse-grained root-mean-square acceleration obtained by averaging the squared modulus of acceleration for all particles located in given boxes of size $\ell$.  Particle voids  correspond clearly to high accelerations, indicating that concentration fluctuations are caused by detachment from the fluid and expulsion from active regions. It is worth noting that this mechanism is distinct from the conventional picture of inertial ejection from simple vortices by centrifugal forces, as the thickness of vortex filaments is several times smaller than the coarse-graining scale $\ell$. 

\begin{figure}
\centering
\subfloat{\includegraphics[width=0.49\columnwidth]{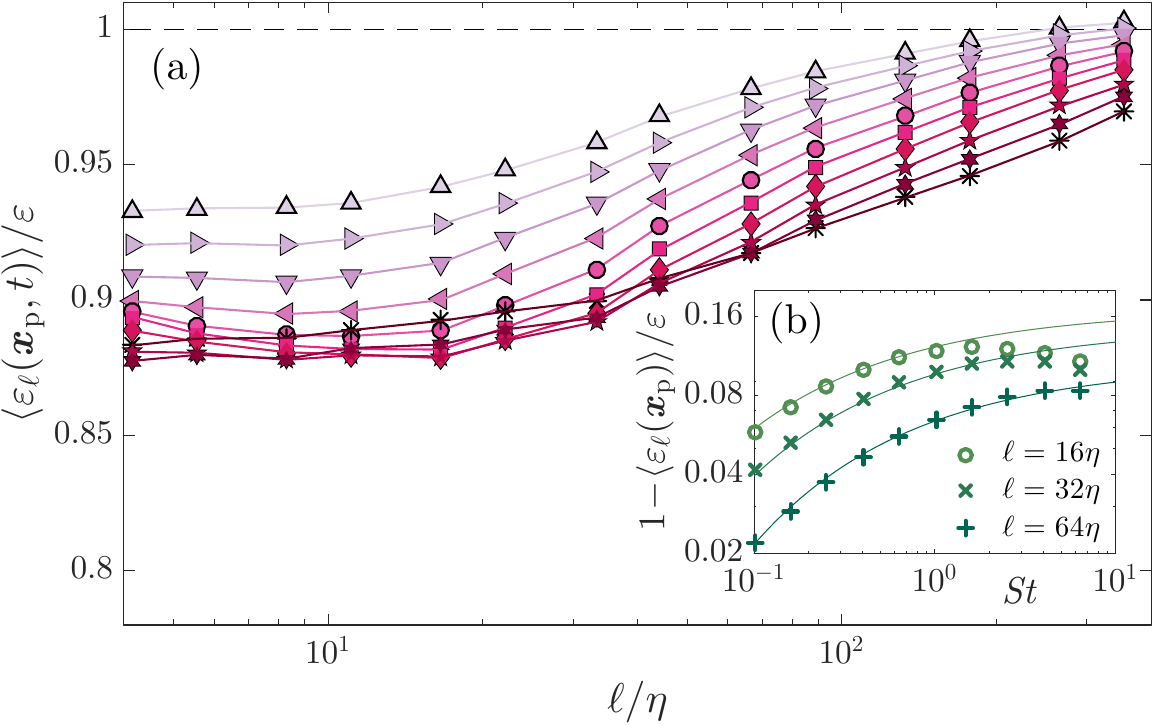}\label{fig:epsil_cg_at_part}}
\subfloat{\label{fig:epsil_cg_at_part_inset}}
\hfill
\subfloat{\includegraphics[width=0.49\columnwidth]{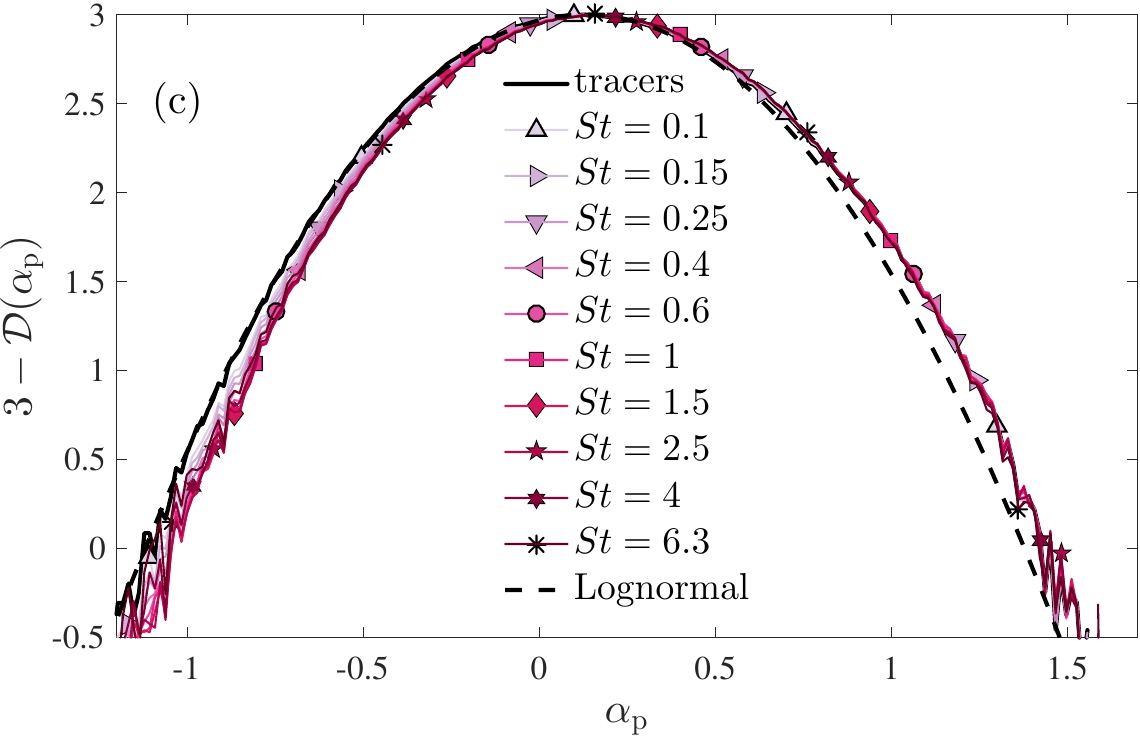}\label{fig:alpha_at_part}}
\caption{(a) Mean coarse-grained dissipation $\epsl$ at particle location as a function of the averaging scale $\ell$ for various Stokes numbers and $R_\lambda = 460$. (b)~Discrepancy shown this time as a function of $\St$ for $\ell=16$, $32$ and $64\eta$. Solid lines are fits $\propto \exp(-c(\ell)/\St^{1/2})$. (c)~Dimension spectrum of the singularity exponent $\alpha_\mathrm{p} = \log (\epsl(\xp)/\avg{\epsl(\xp)}) / \log(\ell/L)$ at particle position for $\ell = 32\,\eta$, various $\St$, tracers, and the lognormal approximation.}
  \label{fig:epsil_at_part}
\end{figure}
Particle concentration in regions of low turbulent activity can be further quantified by measuring the preferential sampling of energy dissipation by the particles. The mean value of $\epsl$ computed along the paths of particles with different Stokes numbers is shown as a function of the coarse-graining scale $\ell$ in figure~\ref{fig:epsil_cg_at_part}. Particles sample preferentially regions where $\epsl$ is lower than the average dissipation $\varepsilon$, even when their inertia is weak --\,see figure~\ref{fig:epsil_cg_at_part_inset}. This bias persists in the inertial range, indicating that it stems from agitation accumulated along particle paths rather than instantaneous ejection from the flow small-scale structures. Measurements of the multifractal spectrum evaluated at particle positions confirm this tendency, as shown in figure~\ref{fig:alpha_at_part} for $\ell=32\,\eta$.  The dependence on $\St$ is weak and visible only at negative values of the singularity exponent corresponding to the most violent events. At positive values of $\alpha_\mathrm{p}$, the dimension spectra associated with different Stokes numbers are almost undistinguishable.  This suggests that preferential sampling results from the expulsion of particles from the most singular regions rather than convergence toward calmer ones.

The observed correlations between the dynamical and concentration properties of particles and the instantaneous inertial-range inhomogeneities of the turbulent flow suggest that the underlying mechanisms are akin to turbophoresis in non-homogeneous flows, at least qualitatively.  Specifically, particles tend to move away from regions with high turbulent activity, forming voids, and follow the fluid in calmer zones.  To provide quantitative support for these ideas, we aim to develop effective equations for an averaged particle density. In the study of turbophoresis in non-homogeneous flows, these equations are obtained by averaging over either the realisations of turbulence or time in statistically stationary and ergodic situations. However, such classical averages are not applicable to instantaneous particle distribution in homogeneous turbulence.  Nevertheless, we expect that a similar effective dynamics can be derived from a low-pass-filtered viewpoint, where the coarse-grained average $\avg{\cdot}_\ell$ plays a central role. 


\section{Non-homogeneous diffusion of Lagrangian trajectories}
\label{sec:lagrangian}

We revisit here the classical approach used to develop stochastic Langevin models for turbulent transport \citep[see, \textit{e.g.},][for a review]{minier2016statistical}. The approach is based on the assumption that while Lagrangian velocities are correlated over timescales of the order of the integral timescale, acceleration become uncorrelated much faster, justifying an approximation of trajectories as diffusive processes. We begin in \S\ref{subsec:fluidacceleration} by providing effective approximations for the second-order statistics of fluid acceleration. We then extend these approximation to inertial particles in \S\ref{subsec:particleacceleration}, specifically to describe their spatially-averaged acceleration. Finally, in \S\ref{subsec:diffusion}, we argue that the coarse-grained dynamics of particles can be approximated as a diffusion process with a space-dependent diffusion coefficient.

\subsection{Fluid acceleration} \label{subsec:fluidacceleration} 
\begin{figure}
\centering
\subfloat{\includegraphics[height=0.31\columnwidth]{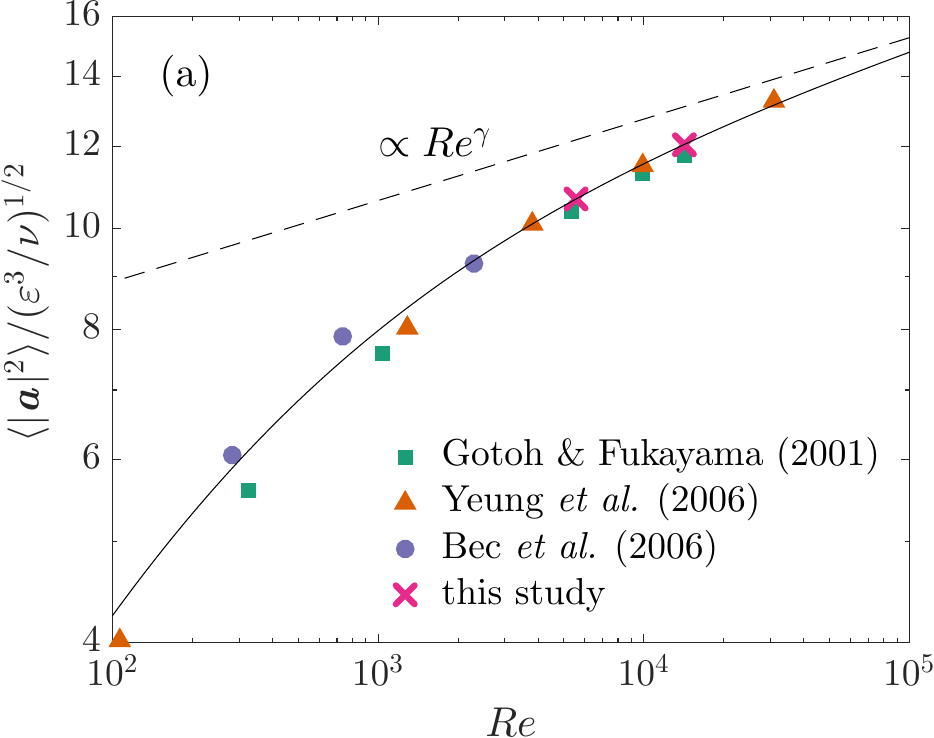}\label{fig:var_accel}}
\hfill
\subfloat{\includegraphics[height=0.31\columnwidth]{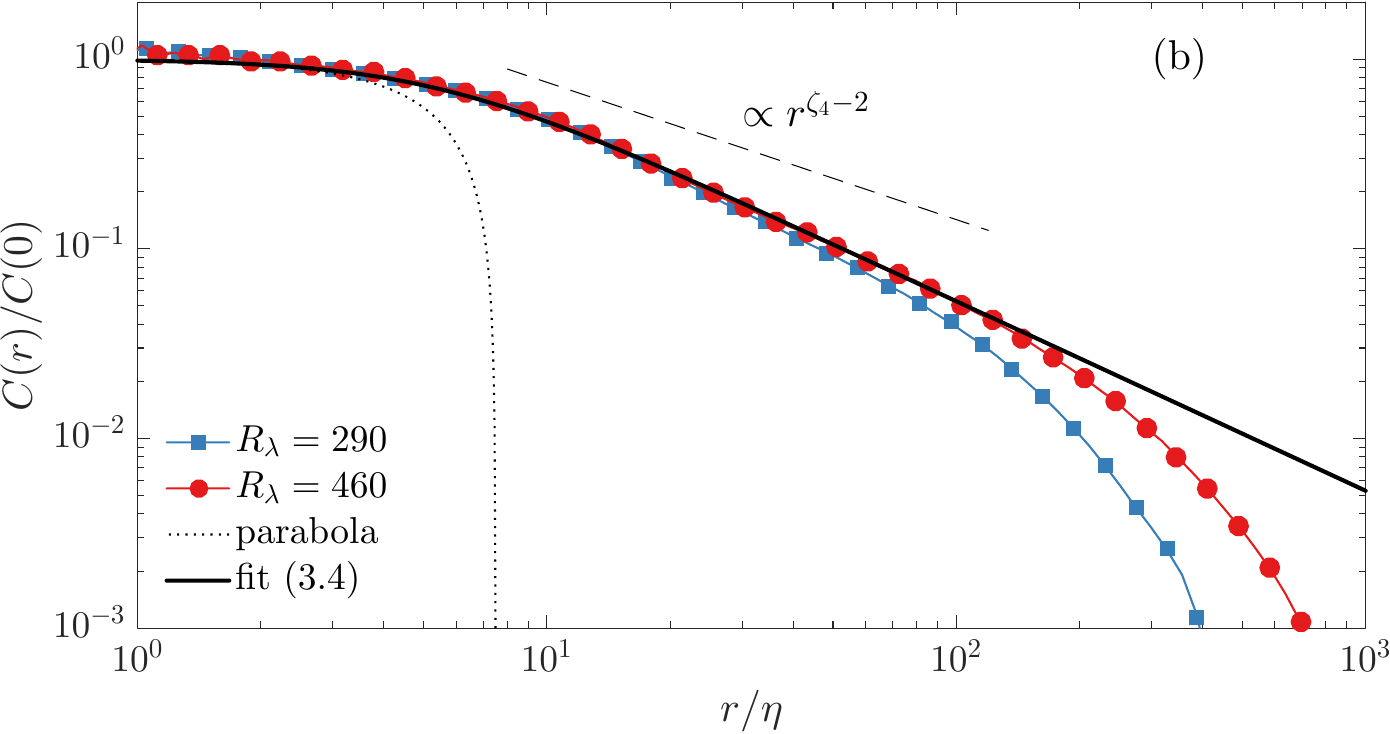}\label{fig:space_corr_accel}}
\caption{(a) Normalised variance of the fluid acceleration as a function of the Reynolds number $\Rey =  R_\lambda^2/15$. Data from several numerical studies are shown as symbols. The dashed line is a behaviour $\propto \Rey^\gamma$ with $\gamma = 0.078$. The solid line corresponds to the fitting formula (\ref{eq:fitA2}) with $a = 6.2$ and $R_\star=80$. (b)~Spatial correlation of the fluid acceleration $C(r) = \langle \ba(\br,t)\boldsymbol{\cdot}\ba(0,t)\rangle$ with $r = |\br|$, shown for the two Reynolds numbers of our dataset. The dashed line is $\propto r^{\zeta_4-2}$ with $\zeta_4 = 1.276$. The dotted curve is the parabolic approximation $C(r) \approx C(0)\,[1-(r/\lambda_1)^2/2]$ with $\lambda_1 = [3\,C(0)/Q(0)]^{1/2} \approx 5.3\,\eta$. The solid line is the approximation (\ref{eq:approx_Correl}) that displays a behaviour $\simeq C(0)\,\lambda_1/r$ at large separations.}
\end{figure}

Turbulent accelerations of fluid particles are one of the most striking signatures of intermittency. At the turn of the century, significant advances in direct numerical simulations  and in particle-tracking experimental techniques have enabled to investigate acceleration statistics in detail \citep[see][for a review]{toschi2009lagrangian}.  These studies revealed that the variance of acceleration deviates from its dimensional estimate and exhibits a notable dependence on Reynolds number.  Specifically, it can be expressed as $\avg{|\ba|^2} = A_2(\Rey)\,\varepsilon^{3/2}/\nu^{1/2}$, where $A_2$ accounts for this dependence.  \cite{hill2002scaling} found that at moderate values of the Reynolds number, Taylor's scaling gives $A_2 \propto \Rey^{1/2}$, assuming that acceleration is dominated by pressure gradients. At large $\Rey$, intermittency prevails and $A_2 \propto\Rey^\gamma$, where $\gamma$ can be estimated using multifractal approaches \citep[see, e.g.,][]{borgas1993multifractal, sawford2003conditional,biferale2004multifractal}.  The log-normal approximation of \S\ref{subsec:turbu} yields $\gamma\approx0.078$ for $\alpha_0= 0.13$.  To match these two asymptotics, we introduce the \textit{ad hoc} approximation:
\begin{equation}
  A_2(\Rey) \approx \frac{a\,\Rey^{\gamma}}{[1+(R_\star/\Rey)^{1/2}]^{1-2\gamma}}.
  \label{eq:fitA2}
\end{equation}
Figure~\ref{fig:var_accel} compares this fit to numerical measurements by \cite{gotoh2001pressure}, \cite{bec2006acceleration}, and \cite{yeung2006acceleration}, together with current simulations.  The approximation~(\ref{eq:fitA2}) with $\gamma=0.078$, $a = 6.2$, and $R_\star=80$ provides a reasonably good agreement.

We now examine the spatial correlations of acceleration, which will be important in approximating particle displacement later. In an isotropic flow, the correlation tensor components in the longitudinal and transverse directions to a given separation $\br$ are interrelated. In a homogeneous and isotropic flow, $C(r) = \langle \ba(\br,t)\boldsymbol{\cdot}\ba(0,t)\rangle$ depends only on the distance $r=|\br|$ and satisfies \citep{obukhov1951microstructure,hill1995pressure} 
\begin{equation}
  \frac{1}{r^2}\frac{\mathrm{d}}{\mathrm{d} r}\left(r^2 \frac{\mathrm{d}C}{\mathrm{d} r} \right) \approx -Q(r), \quad \mbox{with} \ \  Q(r) \equiv \left\langle \partial_ju_i(\br,t)\, \partial_iu_j(\br,t) \, \partial_lu_k(0,t) \, \partial_ku_l(0,t) \right\rangle,
  \label{eq:rel_C_S4}
\end{equation}
where summation is assumed over repeated indices.  This relation assumes that pressure gradients dominate acceleration and uses Poisson equation to express them in terms of velocity gradients.  In homogeneous isotropic flow, the right-hand side of~(\ref{eq:rel_C_S4}) can be expressed as $Q(r) = (1/6)\,\partial_{ijkl} D_{ijkl}(\br) $,  where $D_{ijkl}(\br) = \langle [u_i(\br)-u_i(0)] [u_j(\br)-u_j(0)] [u_k(\br)-u_k(0)] [u_l(\br)-u_l(0)]\rangle$ is the fourth-order structure function. Since $D_{ijkl} \propto r^{\zeta_4}$, we expect acceleration correlations to decrease as $C(r) \propto r^{\zeta_4-2}$ when $\eta\ll r \ll L$.  Therefore, we get $\propto r^{-0.724}$, which is steeper than the K41 prediction $\propto r^{-2/3}$ proposed by \cite{obukhov1951microstructure}. Figure~\ref{fig:space_corr_accel} shows the spatial correlations of acceleration for our two numerical simulations. Our data are consistent with the experimental measurements of \cite{xu2007acceleration} and display a power-law with an even steeper exponent close to $-1$.

This behaviour extends beyond the transition scale introduced by~\cite{hill2002length}, which is derived from the Taylor expansion of correlations at small separations.  Equation (\ref{eq:rel_C_S4}) yields
\begin{equation}
    C(r) = \frac{1}{r}\int_0^r r^{\prime2}\,Q(r^\prime)\,\mathrm{d}r^\prime + \int_r^\infty r^\prime\,Q(r^\prime)\,\mathrm{d}r^\prime.
  \label{eq:C}
\end{equation}
Hence,  the correlation function $C(r)$ can be approximated to leading order as $C(r) \approx  C(0) [1- (1/2)\,(r/\lambda_1)^2]$ as $r$ approaches zero. Here,  $\lambda_1 = [3\,C(0)/Q(0)]^{1/2}$ with $Q(0) =  \left\langle [\mathrm{tr}\, (\boldsymbol{\nabla}\bu)^2 ]^2\right\rangle >0$ is the length scale characterising the parabolic decay of the acceleration correlations, analogous to the Taylor microscale for velocity correlations. It is worth noting that both $C(0)$ and $Q(0)$ exhibit an intermittent dependence on the Reynolds number. While $C(0)$ scales as $\Rey^{\gamma}$ at large Reynolds numbers, $Q(0)$ scales as  $\Rey^{\chi_4}$  \citep[with $\chi_4>0$, see][]{nelkin1990multifractal}. Consequently, we have $\lambda_1 \propto\eta\,\Rey^{(\gamma-\chi_4)/2}$. Using the multifractal lognormal approximation, we obtain $\gamma\approx0.078$ and $\chi_4\approx0.217$, which yields $\lambda_1 \propto\eta\,\Rey^{-0.069}$, consistent with the prediction of \cite{hill2002length}.  Our numerical simulations reveal that $\lambda_1/\eta = 5.51$ for $R_\lambda=290$ and $\lambda_1/\eta = 5.32$ for $R_\lambda=460$, confirming a weak dependence on Reynolds number. Nevertheless, as shown in figure~\ref{fig:space_corr_accel}, deviations from the predicted inertial-range scaling persist for scales much larger than $\lambda_1$.

We interpret the observed behaviour as an extended contribution from small scales to the integral relation~(\ref{eq:C}). For $r>\lambda_1$, the first term always gives a contribution $\simeq Q(0)\,\lambda_1^3/(3\,r) = C(0)\,\lambda_1/r$, obtained by evaluating the integral over the interval $0<r^\prime<\lambda_1$.  Furthermore, separations $r^\prime$ in the inertial range contribute to both integrals a term $\propto\varepsilon^{4/3} r^{-2/3} (r/L)^{\zeta_4-4/3}$, with a universal constant determined by the 4th-order structure function, independent of Reynolds number. Balancing these two terms, we find that the first contribution is dominant as long as $C(0)\,\lambda_1/\eta \gg (\varepsilon^{3/2}/\nu^{1/2})\,(r/\eta)^{\zeta_4-1} (\eta/L)^{\zeta_4-4/3}$, which is satisfied for $r \ll \lambda_2 =\eta\, [(L/\eta)^{4/3-\zeta_4}\,C(0)/(\varepsilon^{3/2}/\nu^{1/2})\,\lambda_1/\eta]^{1/(\zeta_4-1)} \sim \eta\,\Rey^{\alpha}$ where $\alpha = [1-3\,\zeta_4/4+3\,\gamma/2-\chi_4/2]/(\zeta_4-1)$. The lognormal approximation gives $\alpha\approx0.189$, which is smaller than $3/4$, consistently ensuring that $\lambda_2\ll L$. This second crossover scale is much larger than $\lambda_1$, hence ensuring the existence of a range of separations $\lambda_1\ll r \ll \lambda_2$ over which the correlations of acceleration behave as $C(r) \simeq C(0)\,\lambda_1/r$ and this range increases with $\Rey$. The numerical data of figure~\ref{fig:space_corr_accel} confirm this picture. The scaling observed at $r\gtrsim 10\,\eta$ extends further in the inertial range as $\Rey$ increases, and corresponds to $C(r) \propto 1/r$ with a constant that depends weakly on $\Rey$. Both this regime and the small-scale parabolic approximation of the correlation can be matched by the following \textit{ad hoc} formula
\begin{equation}
  C(r) \approx \frac{A_2(\Rey)\,\varepsilon^{3/2}}{\nu^{1/2}[1+(r/\lambda_1)^2]^{1/2}}.
  \label{eq:approx_Correl}
\end{equation}
This approximation, shown as a solid line in figure~\ref{fig:space_corr_accel},  is in good agreement with numerical data. In the following, we will use this formula to coarse-grain the particle dynamics.

\subsection{Particle accelerations}
\label{subsec:particleacceleration}

\begin{figure}
\subfloat{\includegraphics[height=0.31\columnwidth]{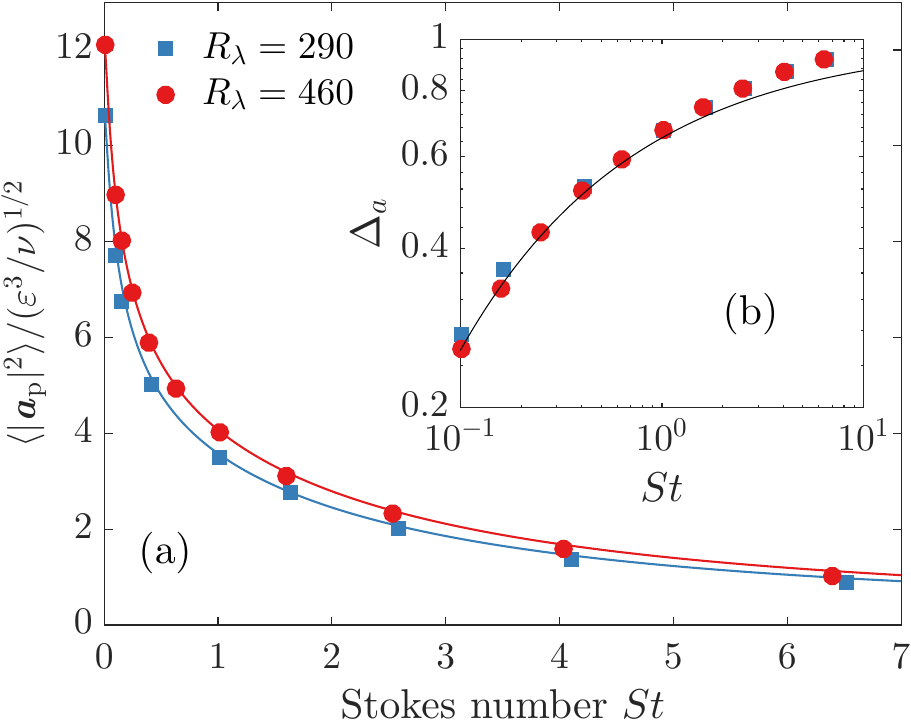}\label{fig:var_accel_part}}
\subfloat{\label{fig:var_accel_part_inset}}
\hfill
\subfloat{\includegraphics[height=0.31\columnwidth]{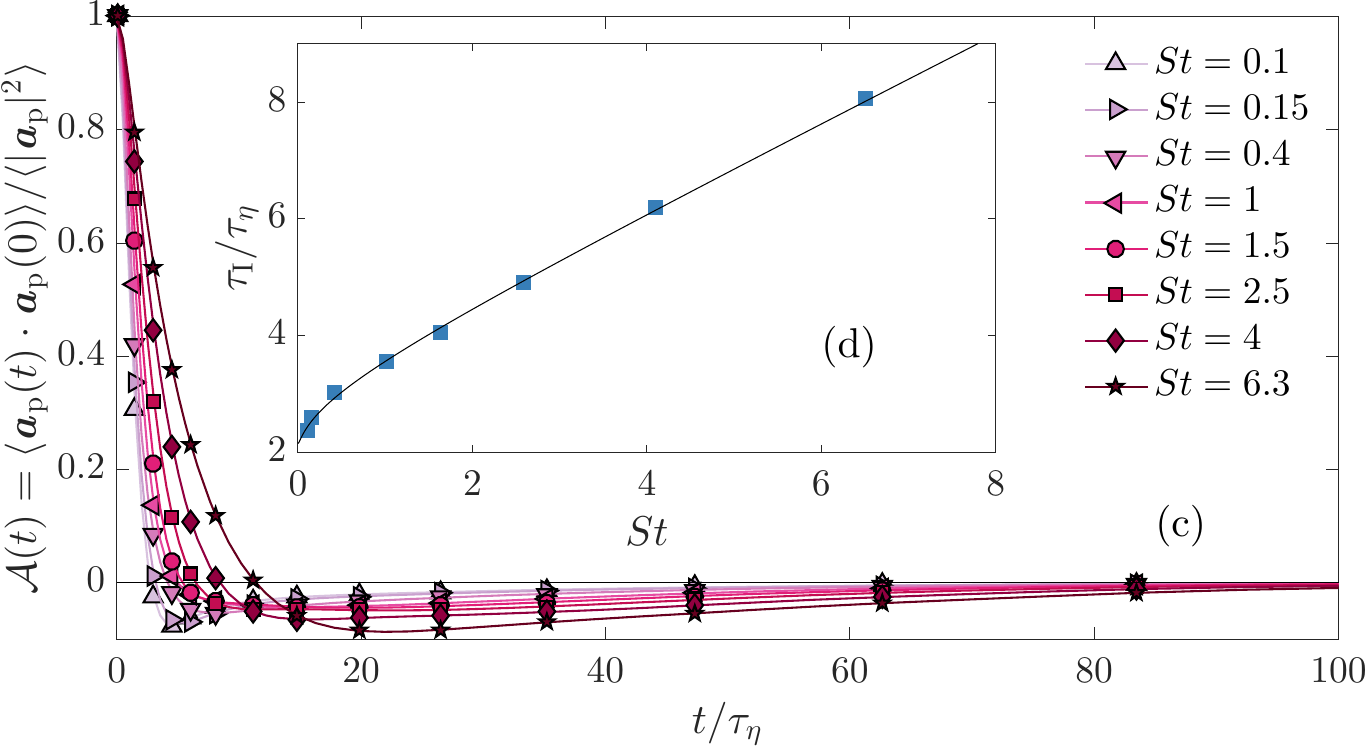}\label{fig:time_correl_accel}}
\subfloat{\label{fig:time_correl_accel_inset}}
  \caption{(a) Variance of particle acceleration for the two Reynolds numbers as a function of the Stokes number.  The two solid curves are corresponding fits of the form (\ref{eq:fitvaraccel}) with $b=0.42$ and $c = 0.17$. (b) Relative discrepancy in acceleration variance in log-log coordinate. The solid curve is $\Delta_a = \exp(-b/\St^{1/2})$.
    (c) Time autocorrelations of particle acceleration for $R_\lambda = 290$ and various $\St$. (d) Integral correlation time of acceleration. The solid line corresponds to (\ref{eq:fit_tcorr_accel}) with $\tau_\mathrm{I}(0) = 2.15\,\tau_\eta$, $b=0.42$, and $d = 0.2$.}
\end{figure}

We focus here on the statistical properties of the acceleration $\ap = \mathrm{d}\vp/\mathrm{d}t$ of inertial particles. Figure~\ref{fig:var_accel_part} shows its variance as a function of the Stokes number. Our measurements agree with those of \cite{bec2006acceleration} and, as they span larger values of the Reynolds numbers, they allow us to substantiate and extend several observations made in that work.

First, we observe that our data, corresponding to two different Reynolds numbers, collapse reasonably well on the top of each other when plotted as a function of $\St$ and rescaled by the acceleration variance of tracers.  This can be seen in figure~\ref{fig:var_accel_part_inset}, which shows the relative discrepancy in acceleration variance $\Delta _a \equiv [\left\langle|\ba|^2\right\rangle - \left\langle|\ap|^2\right\rangle] / \left\langle|\ba|^2\right\rangle$. Although a weak Reynolds-number dependence is noticeable at very small Stokes numbers, one difficulty distinguishes deviations from possible statistical or numerical errors. Therefore, most effects of intermittency are accounted for by the factor $A_2(\Rey)$ introduced in \S\ref{subsec:fluidacceleration}.  This suggests that acceleration variance can be approximated as $\left\langle|\ap|^2\right\rangle \approx  \left\langle|\ba|^2\right\rangle\,[1+\Delta_a(\St)]$, where $\Delta_a$ is a non-dimensional function of the Stokes number with no significant dependence on $\Rey$.

The second observation is an abrupt reduction in the acceleration variance at small but finite values of $\St$. There is a drop of over 25\% from $St=0$ to $St=0.1$, which we interpret as a consequence of preferential sampling, specifically of particle ejection from violent small-scale vortical structures. Our data suggest that the relative discrepancy $\Delta _a$ increases faster than any a power law of $\St$. This is evidenced by its convexity when plotted in log-log in figure~\ref{fig:var_accel_part_inset}, indicating that the acceleration variance may have an essential singularity at $\St=0$. Such a dependence on Stokes number has been observed previously for the rate at which fold caustics occur \citep{wilkinson2006caustic}, a phenomenon also coined sling effect \citep{falkovich2002droplet}. These same events drive the abrupt depletion observed for energy dissipation in \S\ref{subsec:concentrations} and here for acceleration variance.  Figures~\ref{fig:epsil_cg_at_part_inset} and \ref{fig:var_accel_part_inset} show that both discrepancies are well-fitted by a curve $\propto\exp(-c/\St^{1/2})$, where $c$ depends weakly on $\Rey$.  This can be interpreted as a contribution from the probability that the local Stokes number $\taup\,|\boldsymbol{\nabla u}|$ is sufficiently large for the particle to detach from the flow, and thus that $\tau_\eta\,|\boldsymbol{\nabla u}|\gtrsim \St^{-1}$.  At high $\Rey$, the distribution of turbulent velocity gradients is known to display stretched-exponential tails with an exponent $\approx 1/2$ \citep[see][]{yeung2018effects}, consistent with the behaviour of $\Delta_a$. However, \cite{buaria2019extreme} found that the constant in the exponential has a significant dependence on the Reynolds number. Therefore, to further refine our discussion, it will be necessary to better understand this dependence in future studies.

Deviation to this singular behaviour occurs at $\St\gtrsim 1$. Preferential sampling becomes less important, and acceleration statistics are dominated by the particle delay on the flow: their velocity is given by low-pass filtering the fluid velocity over timescales smaller than $\taup$ \citep[see][]{bec2006acceleration}.   \cite{gorokhovski2018modeling} used such considerations to estimate $\left\langle|\bu-\vp|^2\right\rangle \simeq \left\langle|\bu(t)-\bu(t-\taup)|^2\right\rangle \propto \varepsilon\,\taup$, where the last relation assumes $\taup\gg\tau_\eta$ and uses the inertial-range scaling of the second-order Lagrangian structure function.  Consequently, the variance of acceleration approaches a power-law $\left\langle |\ap|^2\right\rangle \propto St^{-1}$ when $\St\gg1$.

The two asymptotics $\St\ll1$ and $\St\gg 1$ can be matched by the \textit{ad hoc} formula 
\begin{equation}
  \avg{|\ap|^2} \approx \frac{A_2(\Rey)\,\varepsilon^{3/2}}{\nu^{1/2}} \,\frac{1-\exp(-b/\St^{1/2})}{(1+c\,\St^2)^{1/4}},
  \label{eq:fitvaraccel}
\end{equation}
which, as seen from figure~\ref{fig:var_accel_part}, gives a fairly good approximation of particle acceleration variance up to $\St\approx 7$. Note that this fitting formula differs from other proposals, such as the one suggested by \cite{gorokhovski2018modeling}, which aimed to also capture response times larger than the large-eddy turnover time $\tau_{L} = u_{\rm rms}^2/\varepsilon$.  As the response times of our particles lie below $\tau_L$ (\textit{i.e.}\/ are such that $\St\ll\Rey^{1/2}$), we use hereafter equation~(\ref{eq:fitvaraccel}). 

We now turn to two-time statistics of the particle acceleration, focusing on the autocorrelation $\mathcal{A}(t) = \avg{\ap(t)\boldsymbol{\cdot}\ap(0)}/\avg{|\ap|^2}$. The results are shown in figure~\ref{fig:time_correl_accel}. Figure~\ref{fig:time_correl_accel_inset} shows the integral time $\tau_\mathrm{I} = \int \mathcal{A}(t)\mathrm{d}t$ as a function of $\St$.  At small Stokes numbers, it approaches the value for tracers, $\tau_\mathrm{I}(0) \approx 2.15\,\tau_\eta$. Deviations occur due again to preferential sampling. Ejection from small-scale vortical structures leads to particles concentrating in regions where the local dissipative timescale is larger than its average.  Dimensionally, we expect $\tau_\eta^{\!@{\rm part}}/\tau_\eta \simeq [\langle|\ba|^2\rangle / \langle|\ap|^2\rangle]^{1/3}$ and, assuming that $\tau_{\rm I} \approx \tau_\eta^{\!@{\rm part}}$, we get $\tau_{\rm I}(\St) \simeq \tau_{\rm I}(0) [ 1-\exp(-b/\St^{1/2})]^{-1/3}$ for $\St\ll 1$. On the other hand, for large Stokes numbers, the particle response time effectively filters out all the flow timescales below it, resulting in $\tau_{\rm I}(\St)\propto\taup$. These two regimes can be matched using the fitting formula
\begin{equation}
  \tau_{\rm I}(\St) = \frac{\tau_{\rm I}(0) [ 1+d\,\St]^{5/6}}{[ 1-\exp(-b/\St^{1/2})]^{1/3}}.
  \label{eq:fit_tcorr_accel}
\end{equation}
where $\tau_\mathrm{I}(0) = 2.15\,\tau_\eta$ is the value measured from tracers, $b=0.42$ is obtained from the acceleration variance, and $d = 0.2$ provides a good agreement with the data of figure~\ref{fig:time_correl_accel_inset}.

To complete this survey, we finally examine the spatially averaged particle acceleration $\avg{\ap}_\ell(\bx,t)$. This quantity is defined as the mean acceleration over all particles that are at time $t$ within a ball $\Bl$ of diameter $\ell$ centred at position $\bx$.  We are particularly interested in the statistics of $\avg{\ap}_\ell$ conditioned on the spatially averaged dissipation rate $\epsl$ obtained in the same ball $\Bl$ at the same time.  This allows us to investigate the relationship between the local fluctuations of small-scale quantities, such as acceleration, and the inertial-range fluctuations of the dissipation field, in accordance with Kolmogorov's refined similarity hypothesis. According to dimensional analysis, the conditional statistics of the coarse-grained acceleration $\avg{\ap}_\ell$, once normalised by $(\epsl^3/\nu)^{1/4}$, depend only on the local Stokes number $\Stl = \taup/(\nu/\epsl)^{1/2}$ and the local Reynolds number $\Rel = \epsl^{1/3}\ell^{4/3}/\nu$.

\begin{figure}
\centering
\subfloat{\includegraphics[width=0.465\columnwidth]{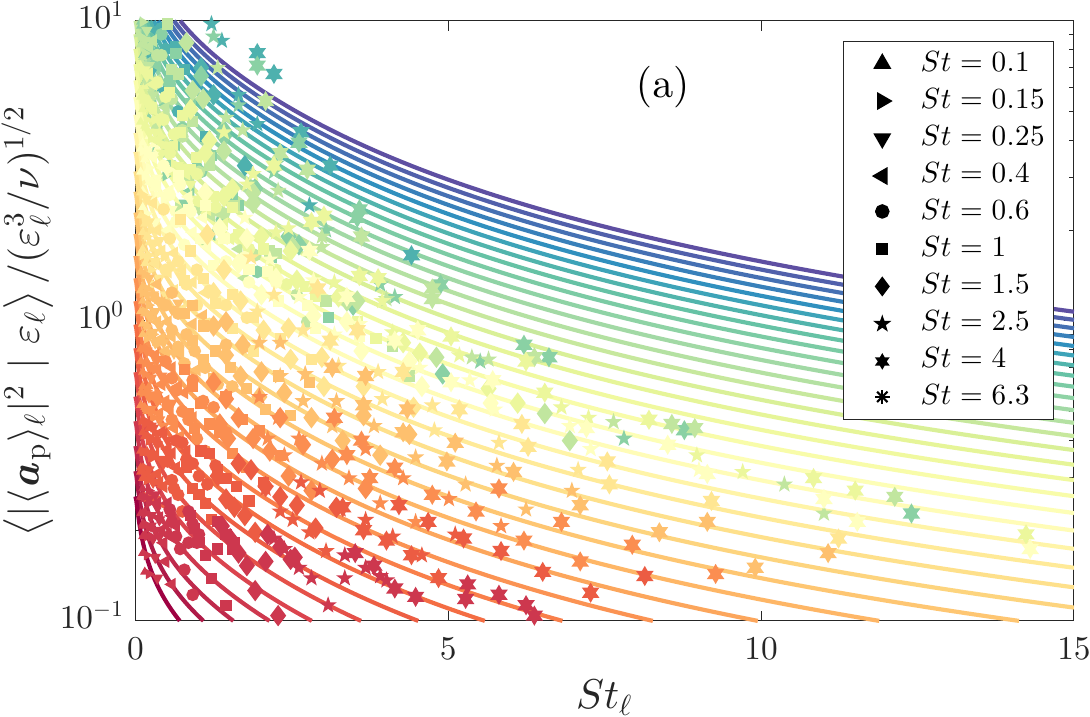}\label{fig:rms_accel_cond_epsil}}
\hfill
\subfloat{\includegraphics[width=0.515\columnwidth]{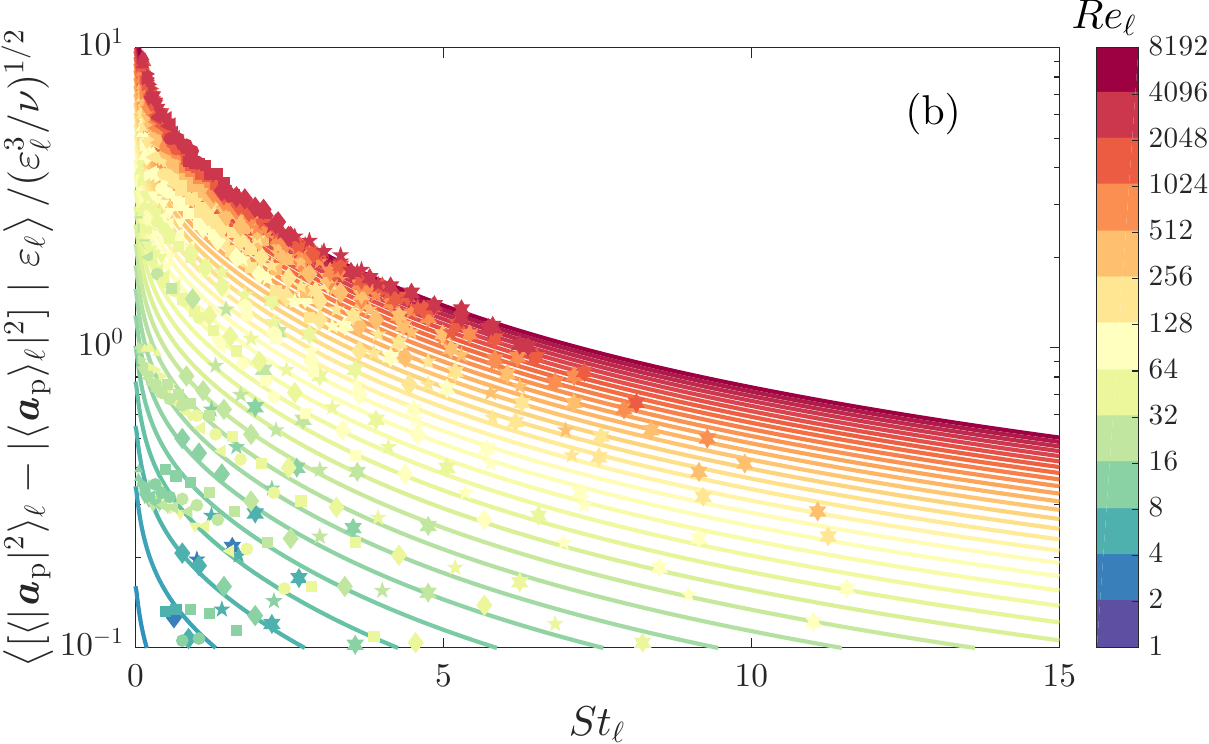}\label{fig:var_accel_cond_epsil}}
\caption{Coarse-grained statistics of particle acceleration conditioned on the local dissipation $\epsl$ for $R_\lambda = 460$. (a) Mean-square coarse-grained acceleration as a function of the local Stokes number $\Stl = \taup/(\nu/\epsl)^{1/2}$ for various $\Rel = \epsl^{1/3}\ell^{4/3}/\nu$ labeled by different colours. Solid lines show prediction (\ref{eq:fit_mean_accel_cond}), while symbols correspond to numerical measurements for different $\St$. (b) Same representation for the coarse-grained variance of particle acceleration. Solid lines are to the fit~(\ref{eq:fit_var_accel_cond}) with $e=2$. }
  \label{fig:var_accel_cond}
\end{figure}
We can express the conditional mean-squared coarse-grained acceleration of particles as
$$
\left\langle |\avg{\ap}_\ell|^2 \ \middle | \ \epsl \right\rangle \propto \frac{1}{\ell^3} \int_{0}^{\ell/2} \left\langle \ap(\br,t)\boldsymbol{\cdot}\ap(0,t) \ \middle | \ \epsl \right\rangle\, r^2 \mathrm{d}r.
$$
Based on our earlier analysis of the spatial correlations of the fluid acceleration and the approximation~(\ref{eq:approx_Correl}), we can assume that for $\Rel\gg1$  $\langle \ap(\br,t)\boldsymbol{\cdot}\ap(0,t) \,|\, \epsl \rangle\approx \langle |\ap|^2  \,|\, \epsl \rangle / [1+(r/\lambda_1(\epsl))^2]^{1/2}$, where $\lambda_1(\epsl) \propto \eta(\epsl)\,\Rel^{(\gamma-\chi_4)/2}$ is the cut-off scale of acceleration spatial correlations associated with the local conditioning dissipation $\epsl$. The conditional acceleration variance $\left\langle |\ap|^2  \ \middle | \ \epsl \right\rangle$ is obtained from (\ref{eq:fitvaraccel}) by replacing $\varepsilon$, $\Rey$ and $\St$ with their local values $\epsl$, $\Rel$ and $\Stl$ given by the spatially-averaged dissipation.  Using $\lambda_1(\epsl) \propto \ell\,\Rel^{-\beta}$ with $\beta = 3/4 + (\chi_4-\gamma)/2 \approx0.819$, we obtain 
\begin{equation}
  \left\langle |\avg{\ap}_\ell|^2 \ \middle | \ \epsl \right\rangle \approx \frac{e}{\Rel^{\beta}} \frac{A_2(\Rel)\,\epsl^{3/2}}{\nu^{1/2}} \,\frac{1-\exp\left(-b/\Stl^{1/2}\right)}{\left(1+c\,\Stl^2\right)^{1/4}},
  \label{eq:fit_mean_accel_cond}
\end{equation}
with $e>0$. This approximation is valid for $\Rel\gg1$, $\ell\ll\lambda_2$, and $\taup\ll\tau_\ell$.  Local fluctuations about this average are described by the coarse-grained variance of acceleration, which we can obtain by replacing the dissipation rate with the conditioning value in~(\ref{eq:fitvaraccel}). We get
\begin{equation}
  \left\langle \avg{|\ap|^2}_\ell-|\avg{\ap}_\ell|^2 \ \middle | \ \epsl \right\rangle\approx \frac{A_2(\Rel)\,\epsl^{3/2}}{\nu^{1/2}} \,\frac{1-\exp\left(-b/\Stl^{1/2}\right)}{\left(1+c\,\Stl^2\right)^{1/4}}\left(1-\frac{e}{\Rel^\beta}\right) .
  \label{eq:fit_var_accel_cond}
\end{equation}

Figure~\ref{fig:var_accel_cond} shows scatter plots of the conditional mean-squared coarse-grained acceleration and the coarse-grained variance of acceleration obtained from numerical simulations. The solid lines in the figure represent the predictions (\ref{eq:fit_mean_accel_cond}) and (\ref{eq:fit_var_accel_cond}), which are based on the approximations made in the preceding text and fitted parameters.  The close agreement between the numerical data and the predictions supports the validity of our approximations.

\subsection{An effective diffusion process}
\label{subsec:diffusion}
To derive effective equations for the particle coarse-grained dynamics, we combine all ingredients from previous analyses. Using equation~(\ref{eq:viscous_drag}), we can write the particle velocity as $\vp(t) = \bu(\xp(t),t) -\taup\,\ap(t)$, allowing us to express its displacement over a time $\delta t$ as
\begin{equation}
  \delta\xp(t) \equiv \xp (t+\delta t)-\xp(t) = \int_t^{t+\delta t}\bu(\xp(s),s)\,\mathrm{d}s -\taup\, \int_t^{t+\delta t}\ap(s)\,\mathrm{d}s.
  \label{eq:dx_1}
\end{equation}
We choose $\delta t$ to be much smaller than the Lagrangian correlation time $\tau_{\rm Lag}$ of $\bu$ to ensure that the fluid velocity along particle path does not vary significantly in $[t,t+\delta t]$. Thus, the first integral in the right-hand side of (\ref{eq:dx_1}) can be approximated as $\bu(\xp(t),t)\, \delta t + O(\delta t/\tau_{\rm Lag})^2$. All fluctuations and dependences on particle inertia are entailed in the second integral. Additionally, if we assume that  $\delta t$ is much longer than the correlation time $\tau_\mathrm{I}(\St)$ of the particle acceleration, we can apply the central-limit theorem and write
$$
\int_t^{t+\delta t}\ap(s)\,\mathrm{d}s \stackrel{\mathrm{law}}{\sim} \mathcal{N}\!\left(\overline{\ap}\, \delta t, \ [\overline{\ap\otimes\ap}-\overline{\ap}\otimes\overline{\ap}]\, \tau_\mathrm{I}\,\delta t\right) + O(\delta t/\tau_{\rm I})^{3/2},
$$
where $\otimes$ is the outer product and $\mathcal{N}\!(\boldsymbol{m}, \mathsfbi{C})$ denotes a multivariate normal random variable with mean $\boldsymbol{m}$ and covariance matrix $\mathsfbi{C}$. The Lagrangian time average $\overline{(\cdot)}$ introduced here is obtained by time integration along particle paths over the interval $[t,t+T]$, assuming the limit $T/\tau_{\rm I}\to\infty$.  It contains information about the turbulent state in which the particle is at the initial time $t$ and is crucial to account for inertial-range fluctuations. To estimate this time average, we use an Eulerian spatial average over a coarse-graining scale $\ell$, so that $\overline{(\cdot)} \simeq \avg{\cdot}_\ell(\xp(t),t)$. This estimate assumes that $T$ is chosen of the order of the turnover time $\tau_\ell = \varepsilon^{-1/3}\ell^{2/3}$ associated with $\ell$, and hence that $\tau_\ell \gg \tau_{\rm I}(\St)$. Preferential sampling by particles, which naturally arises from the Lagrangian average, is now accounted for by evaluating the Eulerian average at the current particle position $\xp$. 

Under these assumptions, we can now express the particle displacement as
\begin{equation}
  \delta\xp(t) \approx \left[\bu(\xp(t),t)-\taup\,\avg{\ap(t)}_\ell\right] \delta t + \boldsymbol{\sigma}_\ell(\xp(t), t)\,\delta \boldsymbol{W}(t).
  \label{eq:dx_2}
\end{equation}
Here, $\delta \boldsymbol{W}$ denotes the increment of the three-dimensional Wiener process, and $\boldsymbol{\sigma}_\ell$ is a tensorial diffusion coefficient that satisfies
\begin{equation}
  \frac{1}{2}\boldsymbol{\sigma}_\ell\,\boldsymbol{\sigma}_\ell^\mathsf{T} = \mathsfbi{D}_\ell \quad\mbox{with}\quad  \mathsfbi{D}_\ell= \frac{1}{2}\taup^2 \tau_\mathrm{I}\, [\avg{\ap\otimes\ap}_\ell-\avg{\ap}_\ell\otimes \avg{\ap}_\ell].
  \label{eq:coeffD}
\end{equation}
This diffusion coefficient not only depends on the particle response time and the coarse-graining scale, but also fluctuates in space and time.  Taking the limit $\delta t \to 0$ while keeping $\tau_\mathrm{I}\ll\delta t\ll \tau_\ell = \varepsilon_\ell^{1/2}\ell^{2/3}$, we can write the effective displacement (\ref{eq:dx_2}) as the stochastic differential equation:
\begin{equation}
  \mathrm{d}\xp(t) \approx \left[\bu(\xp(t),t)-\taup\,\avg{\ap(t)}_\ell\right] \mathrm{d} t +\boldsymbol{\sigma}_\ell(\xp,t)\,\mathrm{d} \boldsymbol{W}(t),
  \label{eq:dx_final}
\end{equation}
where $\boldsymbol{\sigma}_\ell$ is given by (\ref{eq:coeffD}). The diffusion appears here as a multiplicative noise, which we define using the It\^o convention. This is imposed by the requirement that in the statistical steady state, the average particle velocity should vanishes, \textit{i.e.} $\avg{\mathrm{d}\xp(t)/\mathrm{d}t} = 0$. Since $\avg{\bu(\xp(t),t)} = 0$ and $\avg{\avg{\ap}_\ell} = \avg{\avg{\ap}}_\ell = 0$, the contribution of noise should vanish as well.

The proposed model (\ref{eq:dx_final}) for particle dynamics share some similarities with the model introduced by \cite{fevrier2005partitioning}. In both cases, the drift term, the ``mesoscopic Eulerian particle velocity'' in their work, is the sum of the fluid velocity and a residual one. In our model, this residual velocity is proportional to the filtered particle acceleration.  Both models also include a noise term. However, while the ``quasi-Brownian velocity'' of \cite{fevrier2005partitioning} satisfies a molecular chaos assumption and is uncorrelated in space, we identify it in our model as a diffusion with a space-time dependent coefficient that fluctuates due to turbulent agitation. As a result, this contribution is correlated over inertial-range separations.

\begin{figure}
\centering
\subfloat{\includegraphics[height=0.31\columnwidth]{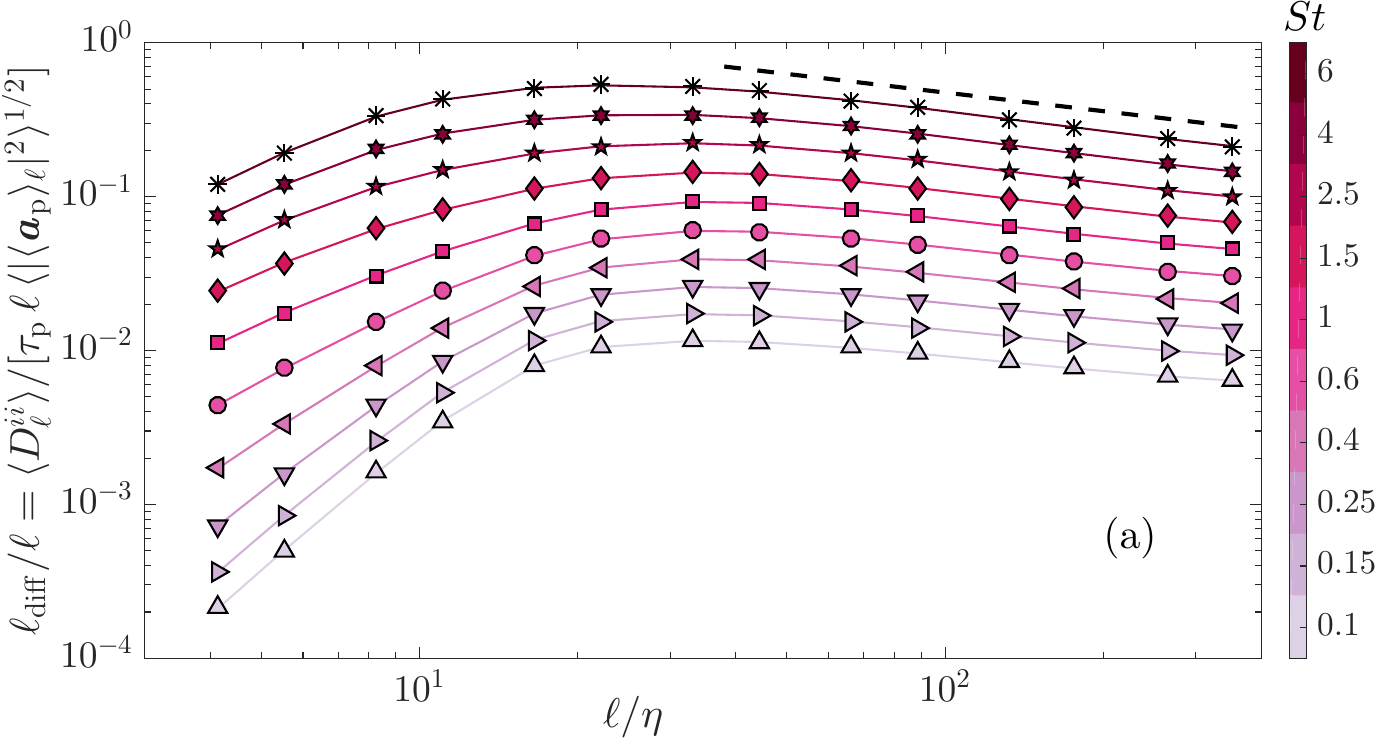}\label{fig:batch_scale}}
\hfill
\subfloat{\includegraphics[height=0.31\columnwidth]{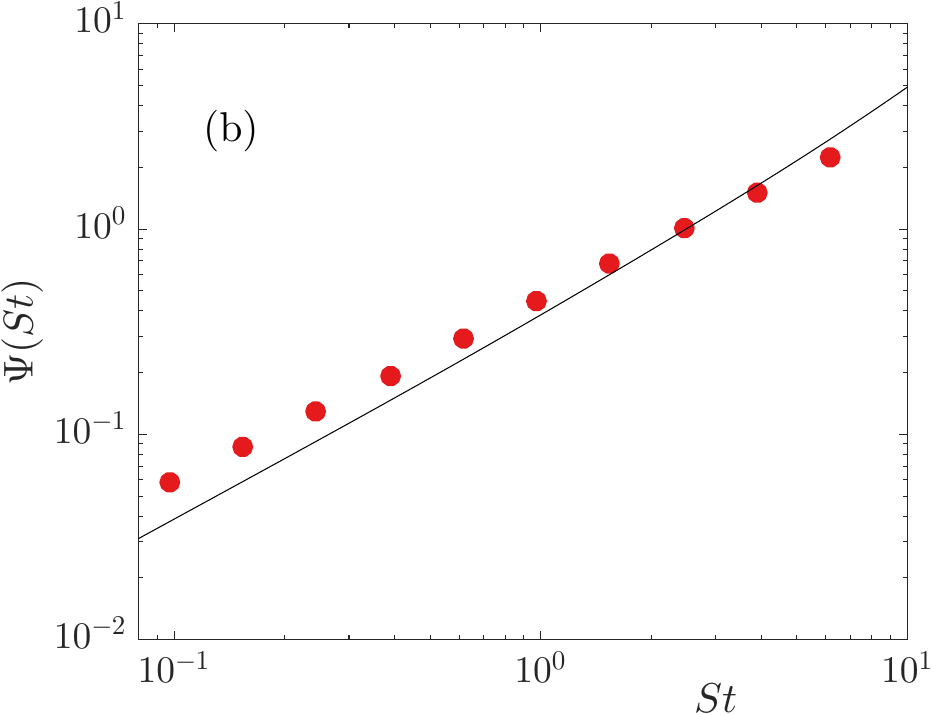}\label{fig:const_ldiff}}
  \caption{(a) Ratio of the diffusive scale $\ell_\mathrm{diff}$ to the coarse-graining scale $\ell$ for various Stokes numbers at $R_\lambda = 460$. The scale  $\ell_\mathrm{diff}$ is the ``Batchelor'' scale above which the diffusive term in (\ref{eq:dx_final}) becomes negligible. The dashed line has a slope $-1+(2/3)(\gamma+\beta) \approx -0.402$, as predicted from the lognormal approximation. (b) Coefficient $\Psi$ of the power law~(\ref{eq:ldiff_predict}). Symbols are numerical measurements and the solid curve is the prediction~(\ref{eq:Psi_predict}).}
  \label{fig:compare_contrib}
\end{figure}

Particle inertia affect both drift and diffusion in the stochastic equation~(\ref{eq:dx_final}). These two contributions have different weights at different scales. They balance each other at a scale $\ell_{\rm diff}$, estimated as $\ell_{\rm diff} = \avg{\mbox{\textit{\textsf{D}}}_\ell^{ii}}/\left[ \taup \avg{|\avg{\ap}_\ell|^2}^{1/2} \right]$. Diffusion dominates at scales smaller than $\ell_{\rm diff}$ and is negligible at larger scales. Thus, the diffusion term is relevant only when $\ell_{\rm diff}$ is larger than the coarse-graining scale $\ell$. Using considerations on acceleration from the previous subsection, we can approximate for $\ell\gg\eta$
\begin{equation}
  \frac{\ell_{\rm diff}}{\ell} = \frac{\taup}{2\,\ell} \frac{\left\langle\tau_{\rm I} \left[ \avg{|\ap|^2}_\ell-|\avg{\ap}_\ell|^2\right]\right\rangle}{\avg{|\avg{\ap}_\ell|^2}^{1/2}} \simeq \Psi(\St)\,\left( \frac{\ell}{\eta}\right)^{-1+(2/3)(\gamma+\beta)}.
  \label{eq:ldiff_predict}
\end{equation}
The exponent is negative, indicating that the diffusive scale becomes very small when the coarse-graining scale $\ell$ is far inside the inertial range.  Numerical measurements of $\ell_{\rm diff}$, reported in figure~\ref{fig:batch_scale}, obtained from the coarse-grained statistics of particle acceleration, confirm the power-law behaviour~(\ref{eq:ldiff_predict}) at $\ell\gg\eta$. We also observe that, for the moderate values of the Stokes number considered, $\ell_{\rm diff}$ is always smaller than $\ell$. Based on previous acceleration correlation measurements, we expect that the constant $\Psi$ behaves as
\begin{equation}
  \Psi(\St) \propto \frac{\St\,(1+d\,\St)^{5/6}\,[1-\exp(-b/\St^{1/2})]^{1/6}}{(1+c\,\St^2)^{1/8}}.
  \label{eq:Psi_predict}
\end{equation}
This prediction, shown as solid curve in figure~\ref{fig:const_ldiff}, compares well with the numerical measurements shown as circles. Extrapolating this behaviour to higher Stokes numbers, we obtain that $\Psi \sim \St^{3/2}$, implying that neglecting diffusion requires choosing a coarse-graining scale such that  $\ell/\eta \gg \St^{(3/2)/[1-(2/3)(\gamma+\beta)]}\approx \St^{3.73}$. Note that this condition applies to particle response times in the inertial range but still smaller than the fluid velocity Lagrangian correlation time $\tau_{\rm Lag}$. This condition is more restrictive than the classical idea that the particle response time should be smaller than the eddy turnover time $\tau_\ell = \varepsilon^{-1/3}\ell^{2/3}$ associated with the coarse-graining scale, which would instead lead to  $\ell/\eta \gg \St^{3/2}$.

In this section, we have shown that the coarse-grained dynamics of inertial particles can be approximated using an effective stochastic equation that includes both drift and diffusion terms. The terms that arise due to particle inertia and account for the differences between the particle and fluid dynamics are governed by the coarse-grained particle acceleration, which fluctuates in both space and time and serves as a clear indicator of turbulent activity. We have also established that for sufficiently large spatial-averaging scale, or equivalently small Stokes numbers, diffusive effects become negligible. In the following section, we will focus on this asymptotics and develop a further level of modelling that will allow us to derive an effective dynamics for the Eulerian coarse-grained density of particles.

\section{Particle transport as an Eulerian ejection process}
\label{sec:eulerian}

\subsection{Model dynamics for the particle density}
\label{subsec:eulerian_model}

\begin{figure}
  \centerline{\includegraphics[width=.9\columnwidth]{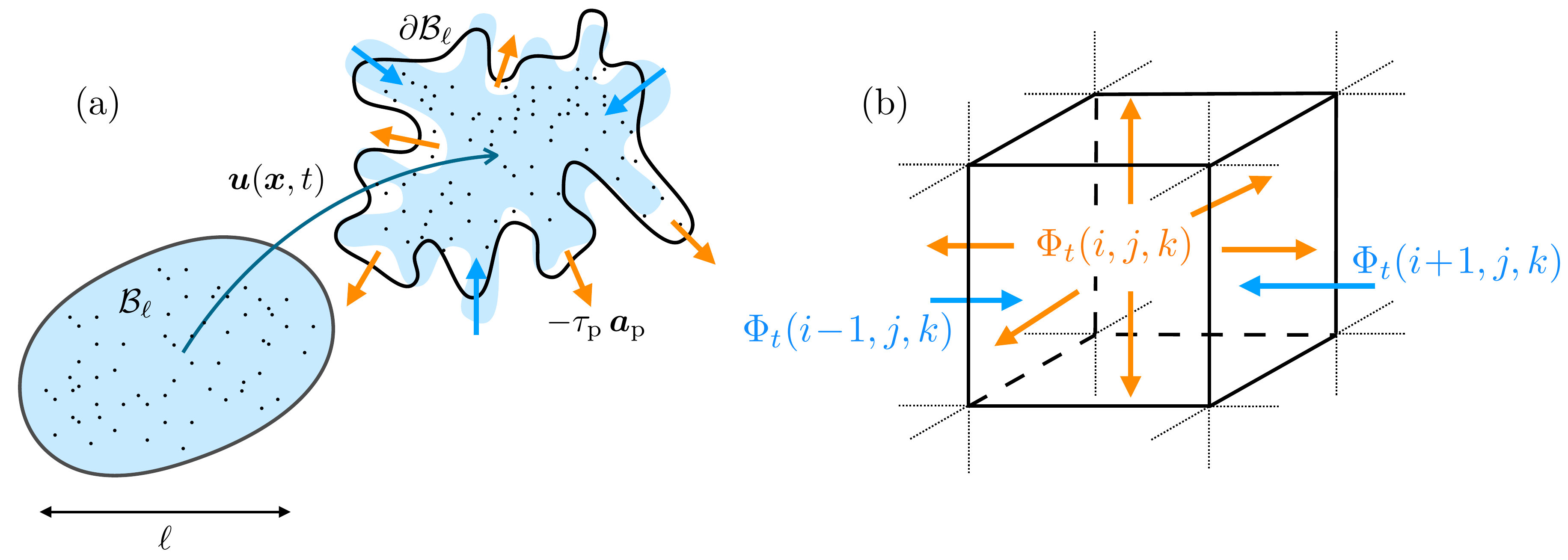}}
  \subfloat{\label{fig:sketch_eject_left}}
  \subfloat{\label{fig:sketch_eject_right}}
  \caption{Sketch of the ejection process. (a) A quasi-Lagrangian viewpoint is adopted to emphasise discrepancies due to inertia. Outgoing fluxes correspond to particles that acquired large-enough accelerations within $\Bl$. (b) The reference cell $(i,j,k)$ ejects particles with a rate $\Phi_t(i,j,k)$ to all its neighbours and receives their individual contributions.}
\end{figure}

In the previous section, we introduced an effective velocity field $\vp^{\rm eff} = \bu - \taup\,\cg{\ap}$, which describes the Lagrangian dynamics of particles for a large enough coarse-graining scale $\ell$ (corresponding to the limit of weak inertia). This approach can be reformulated in a Eulerian frame by considering the evolution of the  coarse-grained particle density $\cg{\rhop}$ in a volume $\Bl$ of size $\ell$ ---\,figure~\ref{fig:sketch_eject_left}. We adopt a quasi-Lagrangian approach and follow the control volume in its motion with the fluid velocity $\bu$, while considering its exchanges with its Eulerian neighbours. To evaluate the fluxes due to particles inertia at the boundary $\partial\Bl$ of the control volume, we distinguish between outgoing and incoming fluxes. Some particles leave the volume because they have acquired a large-enough acceleration inside $\Bl$, and the outgoing flux should thus be controlled by the coarse-grained acceleration $\cg{\ap}$ computed inside the reference volume.  This flux can be expressed as 
\begin{equation}
  \Phi_t(\bx) \approx \int_{\partial\Bl(\bx)} \left\langle\left(-\taup\,\ap \boldsymbol{\cdot n}\right)\,\theta(-\ap \boldsymbol{\cdot n})\, \rhop\right\rangle_\ell \mathrm{d}S \approx 3\,\ell^2\, \taup\,\cg{\rhop}\,|\cg{\ap}|.
  \label{eq:outgoing_flux_phit}
\end{equation}
Here, $\theta$ denotes the Heaviside function, $\boldsymbol{n}$ is the unit vector normal to the surface of $\Bl$, and the average is taken over accelerations satisfying $\ap\boldsymbol{\cdot n} < 0$ to account only for outgoing particles. Assuming isotropic distribution of the outgoing flux, this signed average can be approximated by $(1/2) |\cg{\ap}|$. The control volume $\Bl$ is chosen as a cube with edge length $\ell$, and the spatial domain is tiled by such cubes ---\,figure~\ref{fig:sketch_eject_right}. The time evolution of the mass $\ell^3\cg{\rhop}$ of particles contained in the cell $(i,j,k)$ is then given by 
\begin{eqnarray}
  \frac{\mathrm{D}}{\mathrm{D}t} \left[\ell^3 \cg{\rhop}\right] =&& -\Phi_t(i,j,k) +\frac{1}{6}\big[\Phi_t(i-1,j,k) +\Phi_t(i+1,j,k)+ \Phi_t(i,j-1,k)\nonumber \\
  && +\ \Phi_t(i,j+1,k) +\Phi_t(i,j,k-1) +\Phi_t(i,j,k+1)\big].
  \label{eq:mass_flux}
\end{eqnarray}
$\mathrm{D}/\mathrm{D}t = \partial_t + \bu\boldsymbol{\cdot\nabla}$ denotes here the material derivative along the trajectories of fluid elements.  Mass is lost from the outgoing flux $\Phi_t(i,j,k)$ in the reference cell and gained from the outgoing flux coming from its six neighbours on the cubic tiling.  The right-hand side of equation~(\ref{eq:mass_flux}) corresponds to the discrete Laplacian of the outgoing flux $\Phi_t$.  By considering the mass evolution on scales much larger than the coarse-graining scale $\ell$, we can write a continuous limit which reads
\begin{equation}
  \partial_t \cg{\rhop} + \bu (\bx,t) \boldsymbol{\cdot \nabla}\cg{\rhop} \approx
  \nabla^2 \left[ \kappa_{\ell}(\bx,t)\,\cg{\rhop} \right ],\quad\mbox{with}\ \kappa_\ell  = \taup\,\ell\,|\cg{\ap}|/2.
  \label{eq:model_rhobar}
\end{equation}
The position- and time-dependent coarse-grained diffusion coefficient, $\kappa_{\ell}$, appears inside the Laplacian as expected for an ejection process. This model for particle transport provides a quantitative extension of the  phenomenological ideas proposed in \cite{bec2007toward}. As we will discuss later, the underlying ejection process gives rise to specific features in the probability distribution of the spatially-averaged density, $\cg{\rhop}$.

The diffusive term in equation~(\ref{eq:model_rhobar}) can be expressed as the divergence of the flux vector $\boldsymbol{\varphi}_t = -\kappa_{\ell}\,\boldsymbol{\nabla} \cg{\rhop} -\cg{\rhop}\,\boldsymbol{\nabla} \kappa_{\ell}$, which consists of two distinct contributions. The first corresponds to \textit{osmotic} forces, resulting in classical Fickian diffusion that enhances mixing alongside fluid advection. The second arises from \textit{turbophoretic} forces due to convection by the velocity $\boldsymbol{\nabla} \kappa_{\ell}$, which drive particles from regions with high $\kappa_{\ell}$, characterised by strong particle accelerations and high turbulent activity, to regions with low $\kappa_{\ell}$. The turbophoretic contribution is responsible for the preferential sampling of particles in the inertial-range, as qualitatively discussed in \S\ref{subsec:concentrations}. To determine whether turbophoretic forces are strong enough to induce significant concentration fluctuations and inertial-range voids, we need to compare the magnitudes of the terms in $\boldsymbol{\varphi}_t$. In particular, turbophoretic forces dominate when $|\boldsymbol{\nabla} \kappa_\ell|/\kappa_{\ell}>|\boldsymbol{\nabla}\cg{\rhop}| /\cg{\rhop}$, which implies that the scale of variation of the diffusion coefficient (and thus of the particle acceleration) should be smaller than that of density.

\begin{figure}
\centering
\subfloat{\includegraphics[height=0.31\columnwidth]{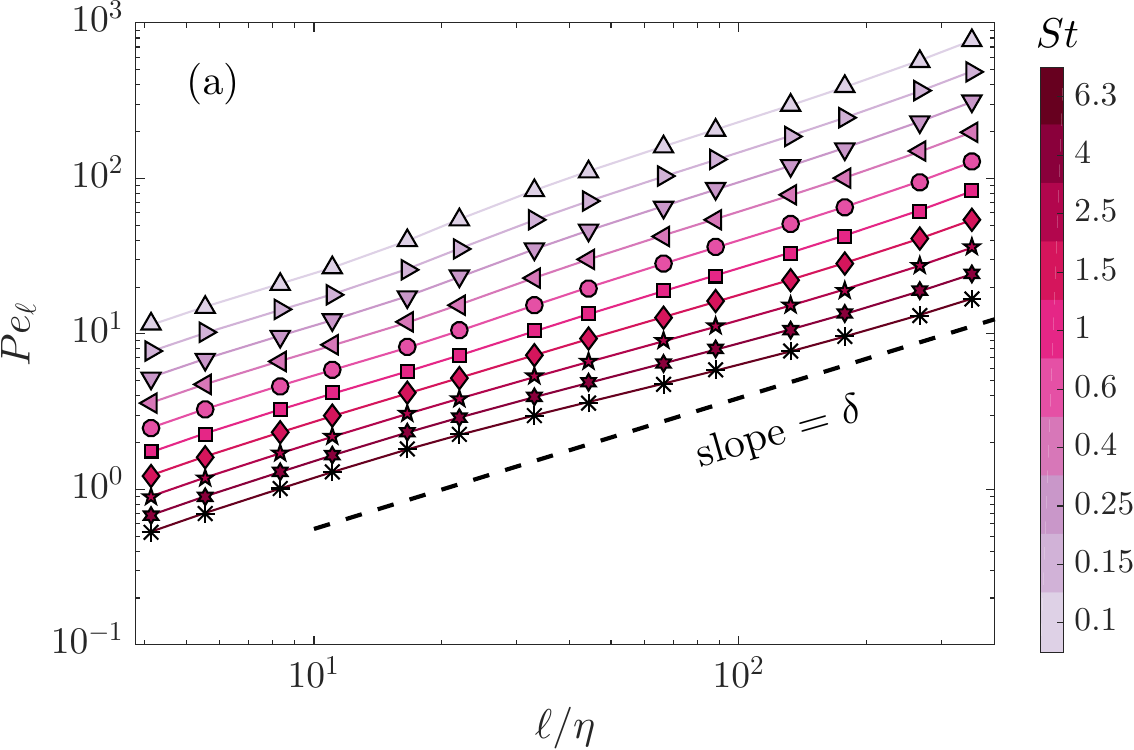}\label{fig:peclet_fn_scale}}
\hfill
\subfloat{\includegraphics[height=0.31\columnwidth]{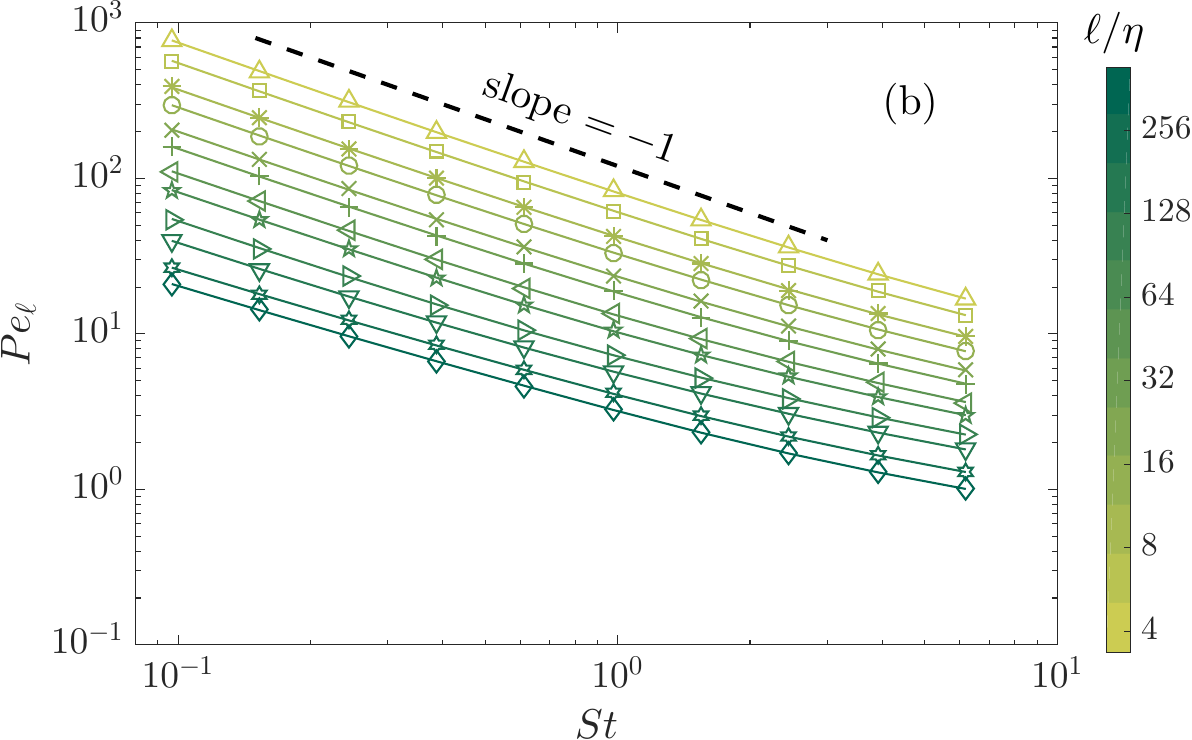}\label{fig:peclet_fn_st}}
\caption{Scale-dependent P{\'{e}}clet number $\Pel$ defined from (\ref{eq:Peclet}) shown in (a) as a function of the coarse-graining scale $\ell$ for various Stokes numbers and in (b) as a function of the Stokes number for various coarse-graining scales $\ell$. The dashed line in (a) shows a behaviour $\Pel \propto (\ell/\eta)^{\delta}$ with $\delta = \zeta_2/2 + (2/3)\,(\gamma-\beta) \approx 0.84$ that is expected for $\ell\gg\eta$. The dashed line in (b) corresponds to $\Pel\propto \St^{-1}$ that prevails at small values of  the Stokes number.}
\end{figure}
The balance between fluid flow convection and turbophoretic diffusion can be characterised at a given coarse-graining scale $\ell$ by a dimensionless P{\'{e}}clet number that we define as
\begin{equation}
  \Pel = \frac{\ell\,\delta u(\ell)}{\avg{\kappa_\ell^2}^{1/2}} = \frac{2\sqrt{3}\,[S_2^{\parallel}(\ell)]^{1/2}}{\taup\, \left\langle|\cg{\ap}|^2\right\rangle^{1/2}}.
  \label{eq:Peclet}
\end{equation}
Here, $\delta u(\ell)$ is the typical fluid velocity fluctuation at scale $\ell$, which is estimated by the square-root of the second-order longitudinal structure function $S_2^{\parallel}$. In the inertial-range, $S_2^{\parallel} \sim \ell^{\zeta_2}$ with $\zeta_2\approx 0.696$. Equation~(\ref{eq:fit_mean_accel_cond}) shows that for the spatially-averaged acceleration, $\langle|\cg{\ap}|^2\rangle\propto A_2(\Rel)\,\Rel^{-\beta}\sim\ell^{(4/3)(\gamma-\beta)}$, which leads to the scaling behaviour $\Pel \sim \ell^{\delta}$ with $\delta = \zeta_2/2 -(2/3)(\gamma-\beta)\approx 0.84$ for coarse-graining scales $\ell$ in the inertial range.  Figure~\ref{fig:peclet_fn_scale} confirms this power-law dependence. Regarding the dependence on the Stokes number, we have $\Pel \sim \St^{-1}$ when $\St\ll 1$, as shown in figure~\ref{fig:peclet_fn_st}. The numerical data indicate that the P{\'{e}}clet number can reach values larger than 1 for both $\ell\gg\eta$ and $\St\ll1$. The scaling laws for small Stokes number and large averaging scale  give $\Pel\sim (\ell/\eta)^\delta/\St$, which results in a P{\'{e}}clet number much higher than unity when $\ell/\eta\gg \St^{1/\delta} \sim \St^{1.19}$.  This scaling is distinct from those discussed in the previous section based on Lagrangian considerations.

\subsection{Distribution of the coarse-grained density}

Based on our previous arguments, we anticipate that for sufficiently large scales, the P{\'{e}}clet number $\Pel$ defined in (\ref{eq:Peclet}) captures alone dependences upon both the Stokes number $\St$ and the coarse-graining scale $\ell$. This asymptotic regime corresponds to the range of parameter values where the approximation~(\ref{eq:model_rhobar}) accurately describes particle dynamics, and we expect that their clustering behaviour will primarily depend on $\Pel$.

\begin{figure}
  \centering
  \subfloat{\includegraphics[width=.49\columnwidth]{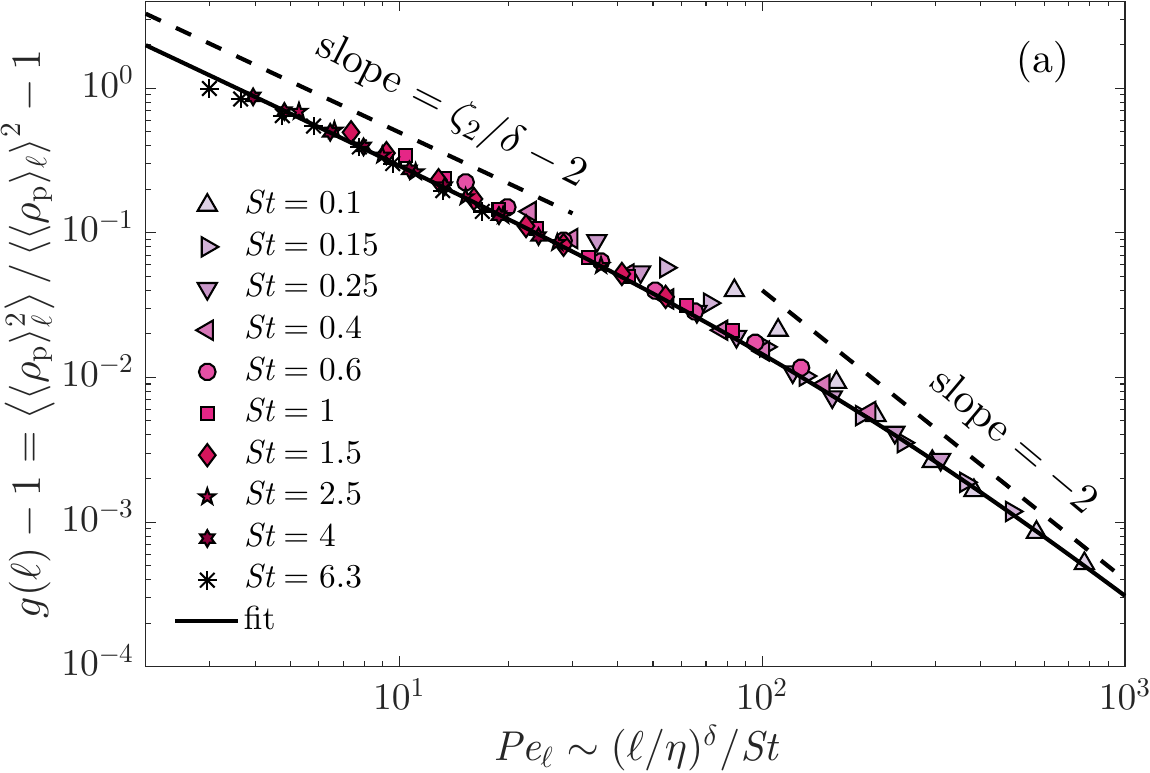}\label{fig:rdf_fn_peclet}}
  \hfill
  \subfloat{\includegraphics[width=.49\columnwidth]{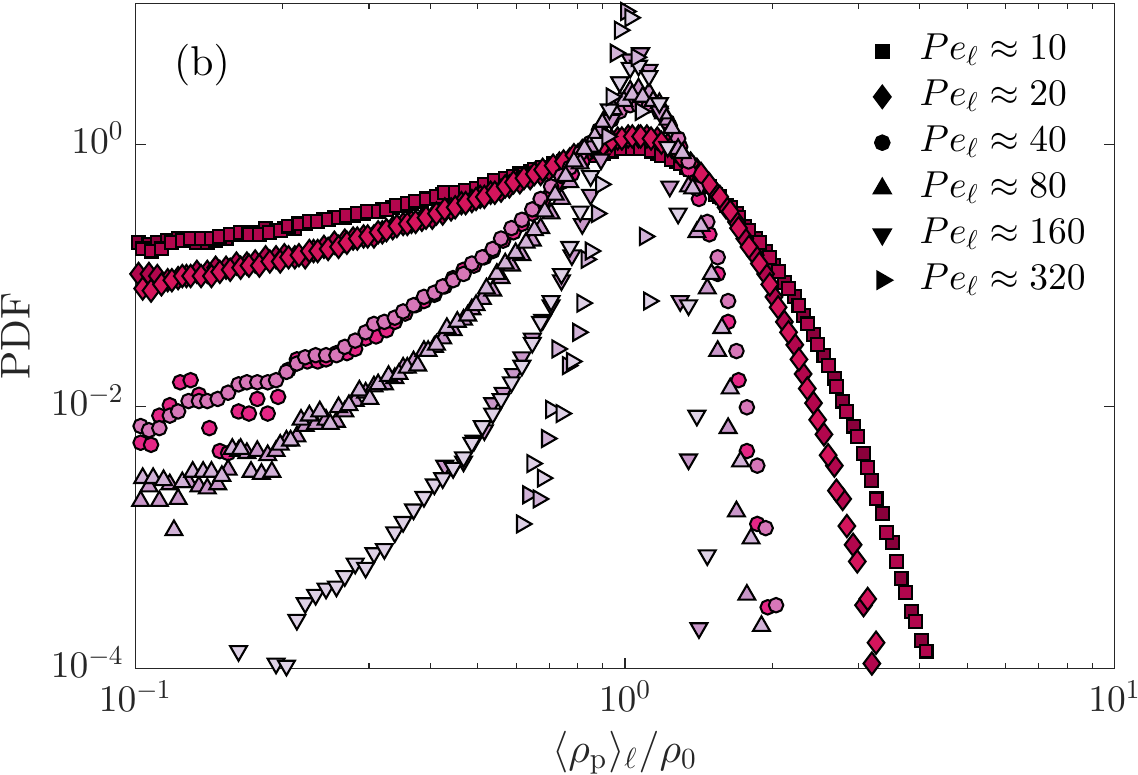}\label{fig:pdf_rho}}
  \caption{(a) Deviation from uniformity of the radial distribution function, $g(\ell)-1$, as a function of the scale-dependent P{\'{e}}clet number $\Pel$ for $R_\lambda=460$ and various Stokes numbers. The dashed lines show the expected behaviours at moderate and large values of $\Pel$. The solid curve is an approximation $\propto \Pel^{\zeta_2/\delta-2} / [1+\Pel/P]^{\zeta_2/\delta}$ with $P = 200$. (b) Probability density function (PDF) of the particle coarse-grained density $\cg{\rhop}$ for six different values of the P{\'{e}}clet number (different symbols), each corresponding to two Stokes numbers (different colours) and two coarse-graining scales combining to the same $\Pel$.}
\end{figure}

We start with examining the radial distribution function, or pair distribution function $g(\ell)$, which describes the probability of finding two particles at a distance $\ell$, normalised by the probability for a uniform distribution. It can be expressed through the second-order moment of the coarse-grained density $\cg{\rhop}$, namely $g(\ell) = \langle\cg{\rhop}^2\rangle / \langle\cg{\rhop}\rangle^2$. For a uniform distribution, we have $\cg{\rhop} \equiv \rho_0 = \langle\cg{\rhop}\rangle$, so $g(\ell)=1$.  Deviations from uniformity as a function of the scale-dependent P{\'{e}}clet number are shown in figure~\ref{fig:rdf_fn_peclet}. Data associated with different values of the Stokes number collapse onto a unique master curve, when the coarse-graining scale $\ell$ is chosen far enough in the inertial range. This curve shows two distinct scaling regimes: one at moderate P{\'{e}}clet numbers and another at large values.

For large values of $\Pel$, we can express the coarse-grained density as $\cg{\rhop} = \rho_0 +\delta\rho$ with $\delta\rho\ll\rho_0$.  To leading order, the perturbation satisfies $\partial_t \delta\rho + \bu\boldsymbol{\cdot\nabla} \delta\rho = \rho_0 \nabla^2 \kappa_{\ell}$.  For statistically stationary deviations to uniformity, we get $\delta u(\ell)\, \delta\rho \sim \rho_0\,\kappa_{\ell} / \ell$, which implies that $\delta\rho\sim\Pel^{-1}$ and the variance scales as $g(\ell)-1 \sim \Pel^{-2}$. At lower P\'{e}clet numbers and higher Stokes numbers, when deviations to uniformity are still small, the velocity contribution is dominated by the large-scale advection, so we have $u_{\rm rms}\,\delta\rho \sim \kappa_\ell/\ell$. This means that deviations from uniformity depend on $\kappa_\ell$, but not on fluid velocity fluctuations at the scale $\ell$.  Thus, we have $\delta\rho \sim \kappa_\ell/\ell \sim \delta u(\ell) /\Pel$. Using $\Pel \sim \ell^{\delta}$, we obtain the second scaling regime $g(\ell)-1 \sim \Pel^{\zeta_2/\delta-2}$.  A solid curve in figure~\ref{fig:rdf_fn_peclet} shows an \textit{ad-hoc} approximation matching these two asymptotic laws.  It provides a reasonable fit to the numerical measurements.

Let us contextualise our results with respect to previous findings on how particles recover a uniform distribution at large scales. When $\ell$ becomes large, $g(\ell)$ tends to unity and in our approach, we find that $\log g(\ell) \approx g(\ell)-1 \sim \Pel^{-2} \sim \taup^2\ell^{-1.68}$. Such an algebraic dependence differs from the exponential decay proposed by \cite{reade2000numerical} that seems confirmed by the experimental measurements of \citet[][see also \citealt{brandt2022particle}]{petersen2019experimental}.
The scaling that we observe also significantly deviates from the prediction of \citet[][see also \citealt{falkovich2003statistics}]{balkovsky2001intermittent}, who proposed that the radial distribution depends primarily on the scale-dependent Stokes number, with  $\log g(\ell) \propto \Stl^2 \sim \taup^2\ell^{-4/3}$ when $\Stl\ll1$.  However, the relevance of $\Stl$ has so far been demonstrated only in models assuming the fluid velocity is a white noise process. For instance, for velocities in the Kraichnan ensemble, it has been shown by \citet{bec2007clustering} that $\log g(\ell) \propto (\mathcal{D}_2(\St_\ell)-3)\log \ell \sim \St_\ell^2\,\log \ell$, which is not a pure function of $\Stl$. Moreover, the direct numerical simulations of \cite{bec2007heavy} at moderate Reynolds numbers suggest that particle distributions primarily depend on a scale-dependent contraction rate $\propto\taup\ell^{-5/3}$, without clear evidence of scaling for the radial distribution function in the inertial range. More recent numerics by \citet{bragg2015mechanisms} and \citet{ariki2018scale} indicate a scaling $\log g(\ell) \sim \ell^{-4/3}$, albeit with uncertainties on the Stokes number dependence.  Our approach reveals a second power-law regime, persisting up to $\Pel\approx100$, where $\log g(\ell) \sim \Pel^{\zeta_2/\delta-2}\sim \tau^{1.17}\ell^{-1.39}$, potentially masquerading a behaviour $\propto \ell^{-4/3}$. It would be of interest to reassess previous measurements of the radial distribution function in the light of present findings.

\begin{figure}
  \centering
  \subfloat{\includegraphics[width=.4\columnwidth]{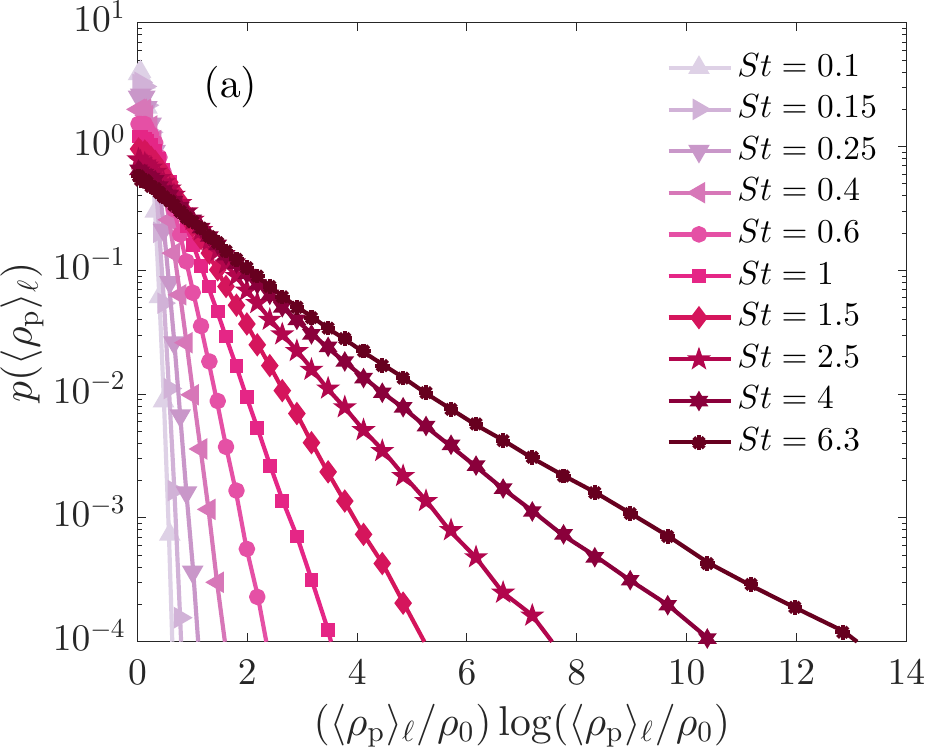}\label{fig:pdf_rho_tail_right}}
  \hfill
  \subfloat{\includegraphics[width=.59\columnwidth]{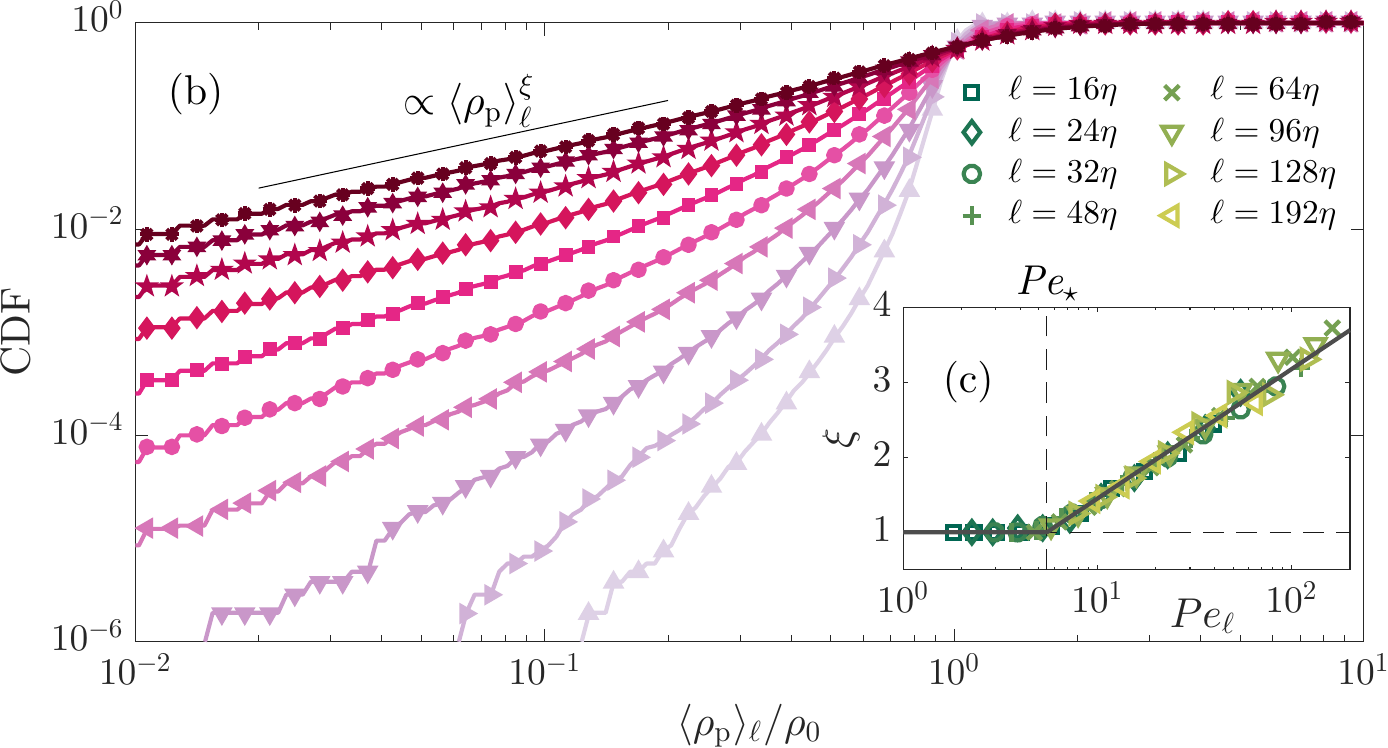}\label{fig:pdf_rho_tail_left}}
  \subfloat{\label{fig:pdf_rho_tail_left_inset}}
  \caption{(a) PDF of the coarse-grained density $p(\cg{\rhop})$ shown for $\ell=64\eta$ represented as a function of $\cg{\rhop}\log \cg{\rhop}$ to evidence the sub-exponential tail at large masses. (b) Cumulative distribution function (CDF) of the coarse-grained density for different Stokes numbers, $\ell=64\eta$ and $R_\lambda=460$ (same labels as panel a), showing a power-law behaviour $\propto \cg{\rhop}^{\xi}$ at small values. Inset (c) Exponent $\xi$ as a function of the scale-dependent P\'{e}clet number, measured for various $\St$ and coarse-graining scales $\ell$.  }
\end{figure}

To complement our analysis, we turn to the probability density function $p(\cg{\rhop})$ of the coarse-grained density. Figure~\ref{fig:pdf_rho} displays numerical measurements for six different high values of the P{\'{e}}clet number. Remarkably, data obtained from various combinations of the particle response time $\taup$ and the coarse-graining scale $\ell$, resulting in the same $\Pel$, exhibit a reasonable collapse, within the range of statistical errors. This confirms the significance of the scale-dependent P{\'{e}}clet number in characterising density fluctuations. The observed probability distributions manifest distinctive features. Both tails, associated with small and large values of $\cg{\rhop}$, are broader than those expected for a Poisson distribution corresponding to a uniform particle density. These deviations can be explained by the ejection process framework developed in \cite{bec2007toward}. Specifically, we find that large densities occur more frequently than the quasi-Gaussian tail of the Poisson distribution. The probability density functions exhibit a sub-exponential behaviour $p(\cg{\rhop})\propto \exp(-C\,\cg{\rhop}\log \cg{\rhop})$, which is clearly captured by our data, as evident in figure~\ref{fig:pdf_rho_tail_right}. 

Regarding the left tail, quasi-empty regions occur also more frequently than in a simple Poisson process. Density distributions follow there a power-law $p(\cg{\rhop})\propto \cg{\rhop}^{\xi-1}$, as depicted in figure~\ref{fig:pdf_rho_tail_left}. This behaviour is again a characteristic feature of ejection processes. To provide a heuristic explanation, we consider the approximation~(\ref{eq:model_rhobar}) of the dynamics, where the evolution of the coarse-grained density along a particle trajectory is given by
$$
\tilde{\rho}_\mathrm{p} \equiv \cg{\rhop}(\xp(t),t) \approx \rho_0 \exp \int_{-\infty}^t \nabla^2\kappa_\ell(\xp(s),s)\mathrm{d}s.
$$
We can thus write the cumulative distribution of the Lagrangian density $\tilde{\rho}_\mathrm{p}$ for $\tilde{\rho}_\mathrm{p}\ll\rho_0$ as
$$
\mathrm{Pr}(\tilde{\rho}_\mathrm{p}<\rho) \approx \mathrm{Pr}\left(\int_{-\infty}^t \nabla^2\kappa_\ell(\xp(s),s)\mathrm{d}s<\log\frac{\rho}{\rho_0}\right) \approx \max_N \left[\mathrm{Pr}\left(\tau_{\ell} \nabla^2\kappa_\ell < \frac{1}{N} \log\frac{\rho}{\rho_0}\right)\right]^N.
$$
Here, we have decomposed the Lagrangian integral of the turbophoretic term into a sum of $N$ equally-distributed independent random variables $\tau_{\ell} \nabla^2\kappa_\ell$, where $\tau_{\ell}$ represents the correlation time of the ejection rate along particle paths. The asymptotic $\rho\ll\rho_0$ behaviour is then obtained by optimising $N$, which represents the number of times mass must be ejected to create a void.  If this number is of the order of unity, the above formula samples the (negative) tail of the distribution of $\nabla^2\kappa_\ell$. A power-law behaviour arises because it is more favourable to choose a value of $N$ of the order of $|\log (\rho/\rho_0)|$, indicating that empty regions are more likely to results from persistent ejections rather than rare, violent events leading to instantaneous voids. Thus, by writing the optimum as $N = -n\,\log (\rho/\rho_0)$ with $n=O(1)$, the cumulative distribution function becomes
$$\mathrm{Pr}(\tilde{\rho}_\mathrm{p}<\rho) \propto \left(\rho/\rho_0\right)^{\tilde{\xi}}, \mbox{ with } \tilde{\xi} = -n\,\log \mathrm{Pr}\left(\tau_{\ell} \nabla^2\kappa_\ell(\xp(t),t) <-1/n\right).$$
When the P\'{e}clet number is large enough, advection dominates, resulting in the correlation time $\tau_\ell$ being given by the eddy-turnover time at scale $\tau_{\ell}\simeq \ell/\delta u(\ell)$. Consequently, $\mathrm{Pr}\left(\tau_{\ell} \nabla^2\kappa_\ell <-1/n\right) \simeq  \mathrm{Pr}\left(|\ell\delta u(\ell)/\kappa_\ell| <n\right) \propto \Pel^{-1}$ for $\Pel\gg1$. Furthermore, the exponent $\tilde{\xi}$ is bounded from below by $0$ in order for the probability distribution of $\tilde{\rho}_{\rm p}$ to be normalisable. Eulerian statistics are then obtained by accounting for the additional factor of $\rho/\rho_0$ involved in the Lagrangian average, because it is itself weighted by the particle density. This finally leads to write the probability distribution of the Eulerian coarse-grained density as
\begin{equation}
\mathrm{Pr}(\cg{\rhop}<\rho) \propto \left(\rho/\rho_0\right)^{\xi}, \mbox{ with } \xi = 1+\tilde{\xi} \approx 1+f\max\left[0,\log(\Pel/\mbox{\it Pe}_\star)\right],
\label{eq:expo_smallmass}
\end{equation}
where $f$ is a positive constant. Figure~\ref{fig:pdf_rho_tail_left_inset} displays the measured exponent $\xi$ as a function of the scale-dependent P{\'{e}}clet number. The exponent saturates at $\xi=1$  for $\Pel<\mbox{\it Pe}_\star\approx 5.5$ and becomes positive above that threshold, increasing as $f\log(\Pel/\mbox{\it Pe}_\star)$ with $f\approx0.75$ for larger values, confirming the prediction given by equation~(\ref{eq:expo_smallmass}).


\subsection{Distribution of voids}
\label{subsec:voids}

We now shift out attention to the large voids that prominently emerge in the spatial distribution of particles. As we observed in \S\ref{subsec:concentrations}, the sizes of these empty regions span the entire inertial range, even at moderate Stokes numbers. Our goal here is to investigate to what extent the statistics of these voids can be explained by the effective diffusion~(\ref{eq:model_rhobar}) introduced in \S\ref{subsec:eulerian_model}.

\begin{figure}
  \centerline{\includegraphics[width=\columnwidth]{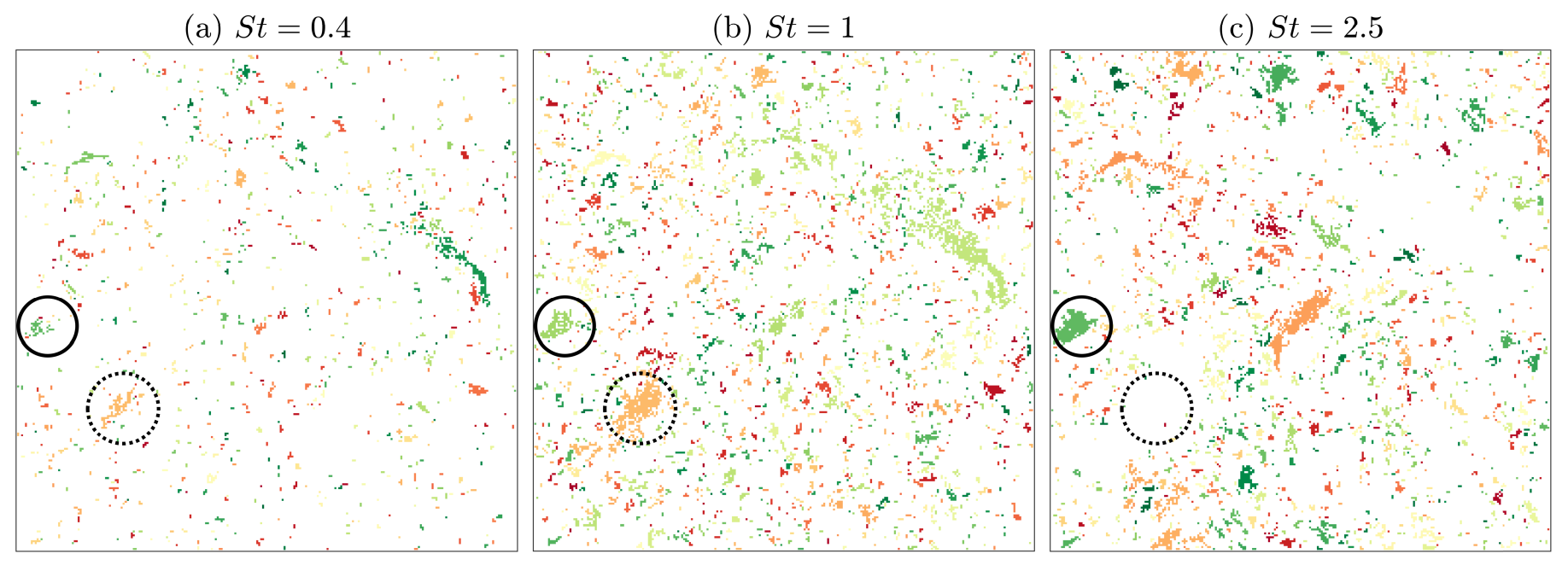}}
  \caption{Two-dimensional cuts of the instantaneous distribution of voids for $R_\lambda=460$ and (a) $\St=0.4$, (b) $\St=1$, and (c) $\St=2.5$. The time and the position of this slice are the same as in figure~\ref{fig:snap_epsil_part}. Voids are obtained as connected empty cubes of size $\ell = 16\,\eta$.}
  \label{fig:snap_voids}
\end{figure}

To detect these voids numerically, we rely on the spatially-averaged density.  They are defined as connected sets of  empty cells, identified by a label-propagation algorithm. The volume $\mathcal{V}$ of each void is determined by counting the number of cubes with a volume $\ell^3$ that it encompasses. While alternative techniques for void detection, such as Delaunay tessellations \citep[see, \textit{e.g.},][]{gaite2005zipf}, may offer better algorithmic efficiency and the ability to define voids in a parameter-free manner, they yield the same results as presented below. Therefore, we have chosen to continue working with the spatially-averaged density, which is central to the model for particle coarse-grained dynamics proposed in \S\ref{subsec:eulerian_model}. Figure~\ref{fig:snap_voids} displays two-dimensional slices of the three-dimensional distribution of voids for a coarse-graining scale of $\ell=16\,\eta$ and for three different values of the particle response time. These distributions are shown at the same instant of time and in the same slice as the local kinetic energy dissipation rate in figure~\ref{fig:snap_epsil_part_a}. The comparison of these two figures clearly identifies voids as regions with high turbulent activity. Furthermore, there are evident correlations between the empty regions associated with different Stokes numbers. One such correlation can be observed for the circled greenish structure, where the intensity of voids increases with $\St$.  Conversely, in other cases, exemplified by the lower orangish structure circled with dots, a void that exists at small $\St$ can be filled by particles with a larger inertia.

\begin{figure}
  \centering
  \subfloat{\includegraphics[width=.49\columnwidth]{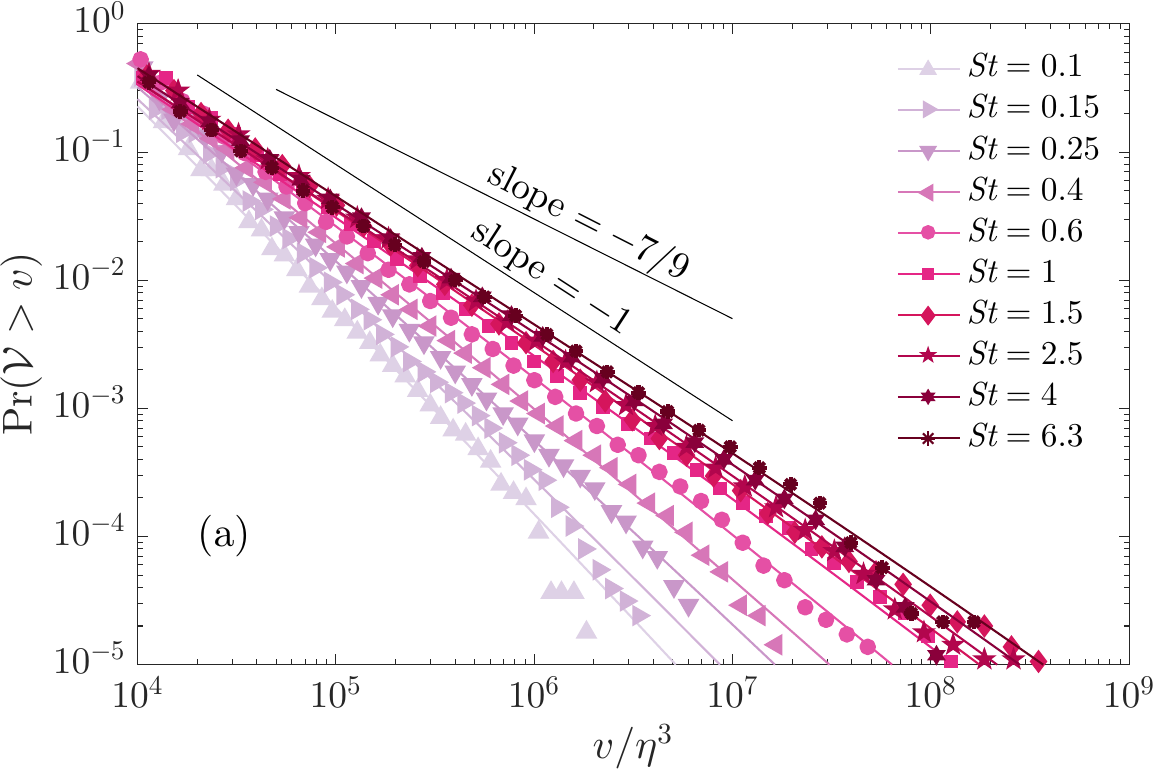}\label{fig:distrib_void_size}}
  \hfill
  \subfloat{\includegraphics[width=.49\columnwidth]{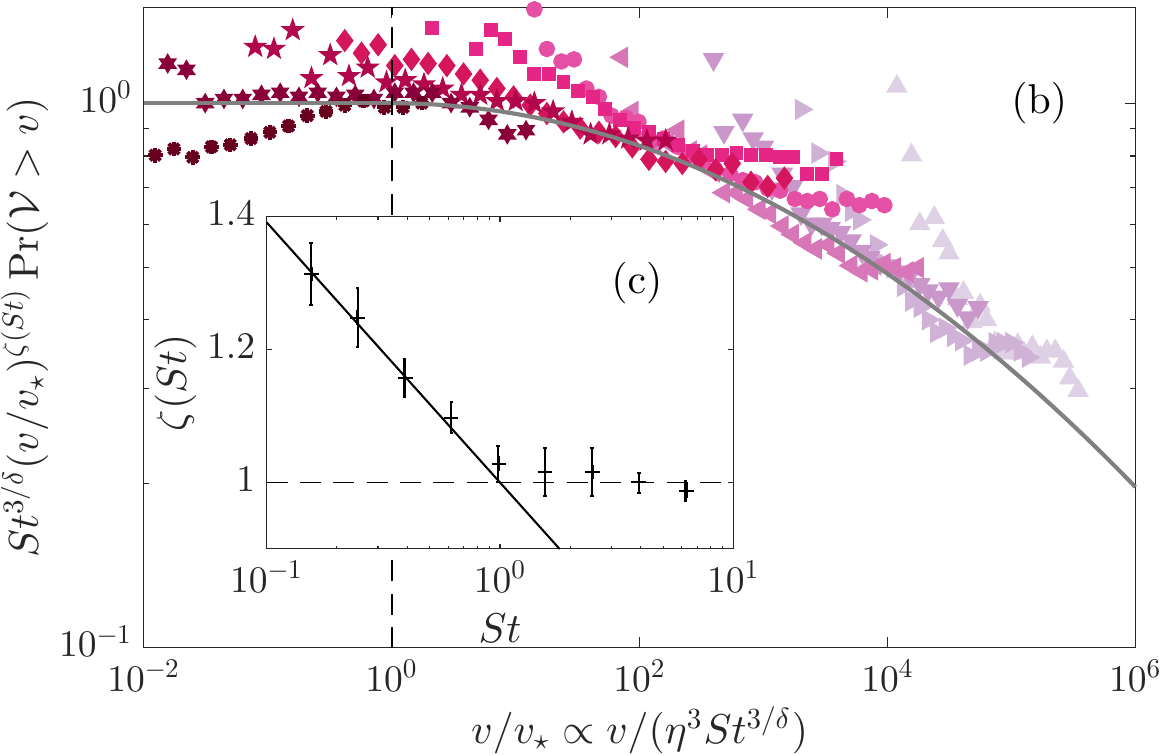}\label{fig:distrib_void_size_rescale}}
  \subfloat{\label{fig:distrib_void_size_rescale_inset}}
  \caption{(a) Complementary cumulative distribution function $\mathrm{Pr}(\mathcal{V}>v)$ of the void volumes $\mathcal{V}$ shown for $R_\lambda=460$ and various Stokes numbers, as labelled. The numerical data is represented by symbols, while the solid lines correspond to approximations of the form~(\ref{eq:CDF_voids}) ---\,see text for parameter values. (b) Same data, rescaled to emphasise the log-normal behaviour for $v\gg v_\star$, shown as a solid line, which is independent of the Stokes number. Inset (c)~Measured power-law exponent $\zeta$ as a function of the Stokes number. The solid line corresponds to $\zeta = 1-0.17\,\log\St$.}
  \label{fig:distrib_voids}
\end{figure}

Figure~\ref{fig:distrib_void_size} presents the complementary cumulative probability distributions of void volumes obtained for different Stokes numbers and an elementary coarse-graining scale $\ell=16\eta$.  These distributions exhibit broad tails at large inertial-range volumes, displaying a distinctive power-law behaviour $\mathrm{Pr}(\mathcal{V}>v) \propto v^{-\zeta}$, where $\zeta\approx 1$ is particularly evident for the highest Stokes number values. These statistics remain robust when using alternative definitions of voids, or when changing either the total number of particles $N_{\rm p}$ or the coarse-graining scale $\ell$. Similar power-law dependencies have been previously observed in the probability distribution of void sizes. In the two-dimensional inverse cascade, \cite{boffetta2004large} found an intermediate regime where the probability density function of void areas behaves as $p(a)\propto a^{-1.8}$, independent of the Stokes number, with an exponential cutoff at larger sizes. \cite{goto2006self} proposed a self-similar distribution of void areas with $p(a)\propto a^{-5/3}$, arising from sweep-stick mechanisms where particles preferentially trace fluid zero-acceleration points. Extending these arguments to three dimensions, \cite{yoshimoto2007self} predicted a power-law exponent $\zeta = 7/9\approx0.778$ for the cumulative distribution of void volumes, with reasonable numerical support at moderate values of the Reynolds number. Figure~\ref{fig:distrib_void_size} showcases this behaviour for comparison. Additional evidence supporting this shallow trend comes from grid-turbulence experiments by \citet{sumbekova2017preferential} and analyses employing Vorono\"i diagrams, where they found $p(a)\propto a^{-1.8\pm0.1}$ for void areas in two-dimensional cross-sections of the three-dimensional particle distribution. Assuming a relationship of the form $p(v) \sim v^{-1/3} p(a)$ with $a\sim v^{2/3}$, these observations suggest $\zeta\approx0.53\pm0.07$. However, our data clearly show a steeper slope, even for Stokes numbers exceeding those considered in both \citet{yoshimoto2007self} and \citet{sumbekova2017preferential}. 

We revisit here void statistics in light of the ejection process that we introduced to model particle dynamics in the inertial range.  The probability that the volume of a void exceeds the value $v$ can be estimated as the probability of finding very few particles in a cube of size $\ell\sim v^{1/3}$. This implies that the coarse-grained density is there of the order of or smaller than $v^{-1}$. Thus, we can write $\mathrm{Pr}(\mathcal{V}>v) \sim \mathrm{Pr}(\cg{\rhop}\lesssim v^{-1})$. Using the asymptotic behaviour~(\ref{eq:expo_smallmass}) for the distribution of $\cg{\rhop}$ at small values, we obtain
$$
\mathrm{Pr}(\mathcal{V}>v) \sim (v/\eta^3)^{-\xi({\scriptsize\Pel})} \sim (v/\eta^3)^{-1}\mathrm{e}^{-\log(v/\eta^3)\,f\max\left[0,\log({\scriptsize\Pel}/{\scriptsize\mbox{\it Pe}}_\star)\right]}.
$$
Choosing $\ell$ to be of the order of $v^{1/3}$, we have $\Pel\sim(\ell/\eta)^\delta/\St\propto(v/\eta^3)^{\delta/3}/\St$, resulting in
\begin{equation}
\mathrm{Pr}(\mathcal{V}>v) \sim \begin{cases} \St^{-3/\delta}(v/v_\star)^{-1} & \mbox{if } v<v_\star, \\
 \St^{-3/\delta}(v/v_\star)^{-\zeta(\St)}\mathrm{e}^{-g\left[\log(v/v_\star)\right]^2}& \mbox{if } v\ge v_\star.\end{cases}
\label{eq:CDF_voids}
\end{equation}
Here, $g$ is a positive constant, $v_\star\propto\eta^3\St^{3/\delta}$, and the exponent $\zeta$ has a logarithmic dependence on the Stokes number of the form $\zeta \approx 1-h\,\log(\St/\St_\star)$, where $h>0$. Figure~\ref{fig:distrib_void_size} shows such predictions for the distribution of void volumes along with numerical data. Reasonable agreement is obtained by choosing for fitting parameters $v_\star/\eta^3=4000\,\St^{3/\delta}$, $g=0.0085$, $h=0.17$, and $\St_\star=1$. The measurements shown in the right-hand panel of figure~\ref{fig:distrib_voids} corroborate these values. Figure~\ref{fig:distrib_void_size_rescale} represents the rescaled complementary cumulative distribution of void volumes as a function of $v/v\star$. Despite statistical noise, data associated with various Stokes numbers (symbols) seem to collapse for $v>v_\star$ onto the log-normal master curve $\exp[-0.0085[\log(v/v\star)]^2]$ (solid line). The measured exponent $\zeta$ is represented in figure~\ref{fig:distrib_void_size_rescale_inset}. It follows  $\zeta \approx 1-0.17\,\log\St$ for $\St\lesssim 1$ and saturates to $\zeta \approx 1$ for larger $\St$.

It is worth noting that the intermediate power-law behaviour that we observe in the distribution of void sizes can be interpreted in terms of Zipf's law~\citep[see, \textit{e.g.},][]{cristelli2012there}.  Samples following this law exhibit coherence and adhere to certain dynamical constraints, which are satisfied when the size dynamics of the objects under consideration can be described as a multiplicative process. In the context of turbophoresis, interpreted as an ejection process, this framework naturally emerges, as the mass of particles ejected from a given cell is proportional to its volume. For such a coherent process, the exponent $\zeta=1$ represents a classical case.  It arises when large voids are formed through the merging of smaller, independent voids with uncorrelated histories, as may occur at large Stokes numbers. The growth rate of a large void becomes proportional to the probability of intersecting other empty regions, which, in turn, is proportional to its volume. This process, known as ``preferential attachment'', leads to an exponent of $\zeta=1$ \citep[see][]{de2021dynamical}.

\section{Concluding remarks}
\label{sec:conclusion}

In this paper, we have presented convincing evidence that the phenomenon of turbophoresis,  previously thought to occur only in turbulent flows containing inhomogeneities, also manifests in statistically homogeneous situations.  This effect arises from the instantaneous non-uniformities intrinsic to turbulent flows, spanning the whole inertial range. Our direct numerical simulations clearly illustrate the ejection of inertial particles from highly active regions of the flow, leading to their concentration in calmer regions. Remarkably, this behaviour persists in spatially coarse-grained representations of both the flow and the particles, resulting in strong correlations between the spatially-averaged particle concentration and the fluctuations in turbulent kinetic energy dissipation within the inertial range.

The fluctuations in particle acceleration play a crucial role in the turbophoresis process. When particles experience pure Stokes drag, these acceleration fluctuations govern their deviations from fluid motion. Through analytical and phenomenological arguments, as well as a detailed analysis of numerical simulations, we have gained insights into the statistics of particle acceleration. This includes understanding spatial and temporal correlations, as well as the influence of fluid flow intermittency on second-order statistics. Building upon these insights, we have introduced approximations for the inertial-range dynamics of particles in terms of effective diffusion equations with a diffusivity that varies in both space and time. The diffusion coefficient is expressed in terms of the coarse-grained particle acceleration, which, in turn, is determined by local turbulent activity. These approximations hold when spatial averaging scales are sufficiently large or particle inertia is sufficiently small, ensuring that higher-order corrections to this dynamics remain negligible. In this asymptotic regime, the dynamics of particles depend solely on a local P{\'{e}}clet number that quantifies the relative importance of advection by the fluid flow compared to inertia-induced diffusion at a given coarse-graining scale $\ell$. Notably this P{\'{e}}clet number exhibits a non-trivial power-law dependence on the observation scale, $\Pel\sim(\ell/\eta)^{0.84}/\St$, where the exponent is prescribed by the intermittent statistics of the fluid velocity and deviates significantly from the value $2/3$ that would be obtained by dimensional analysis.

The diffusive models we have developed provide means to infer of the distribution of particles in the inertial range. Specifically, we demonstrate that the statistics of the coarse-grained particle density $\cg{\rhop}$ at a given inertial-range scale $\ell\gg\eta$ depend solely on the scale-dependent P{\'{e}}clet number $\Pel$.  Furthermore, these diffusive models predict that the probability density functions of $\cg{\rhop}$ exhibit algebraic tails at small values and allow for the characterisation of the associated exponent as a function of $\Pel$. For large masses, the models predict a super-exponential behaviour that is also well reproduced by our direct numerical simulations. However, statistics that span different scales, such as the distribution of voids, display more intricate dependencies. Nonetheless, we find that the probability distribution of void volumes follows a power law with exponent steeper than $-2$ at intermediate values, transitioning to a log-normal tail at larger values. Our direct numerical simulations demonstrate a reasonably good agreement with this prediction, emphasising the need to revisit previous work on void statistics in the light of these potential behaviours.

The introduction of space-dependent diffusions in this study presents a novel framework for incorporating inertial particles into models or large-eddy simulations of turbulent flows. The coarse-grained particle density can be effectively approximated using diffusion equations derived from spatial averaging, with a fluctuating diffusion coefficient determined by the local turbulent dissipation rate. To test, calibrate, and validate this approach, further  numerical simulations that integrate the effective advection-diffusion equations at various coarse-graining resolutions are necessary. Although beyond the scope of this work, this perspective holds promise for future work.


\backsection[Funding]{Computational resources were provided by GENCI (grant IDRIS 2019-A0062A10800) and by the OPAL infrastructure from Universit\'e C\^ote d'Azur. This work received support from the UCA-JEDI Future Investments, funded by the French government (grant no. ANR-15-IDEX-01), and from the Agence Nationale de la Recherche (grant no. ANR-21-CE30-0040-01).}

\bibliographystyle{jfm}

\end{document}